\documentclass[10pt,a4paper]{article}
\usepackage{amssymb,amsfonts,psfrag}
\usepackage{amscd}
\usepackage{amsmath}
\newcommand{\lbl}[1]{\label{eq:#1}}
\newcommand{\rf}[1]{(\ref{eq:#1})}
\newcommand{\nn}{\nonumber}

\newcommand{\be}{\vs{2}\begin{equation}}
\newcommand{\ee}{\vs{2}\end{equation}}

\newcommand{\bea}{\begin{eqnarray}}
\newcommand{\ena}{\end{eqnarray}}
\newcommand{\nnbea}{\begin{eqnarray*}}
\newcommand{\nnena}{\end{eqnarray*}}

\newcommand{\lra}{\ \longrightarrow\ }

\newcommand{\cR}{{\cal R }}
\newcommand{\cW}{{\cal W }}

\newcommand{\cF}{{\cal F }}
\newcommand{\ccC}{{\cal{C} }}

\newcommand{\cN}{{\cal N }}
\newcommand{\cA}{{\cal A }}

\newcommand{\cD}{{\cal D }}
\newcommand{\cL}{{\cal L }}
\newcommand{\ccA}{{\cal A }}

\newcommand{\ccM}{{\cal{M} }}
\newcommand{\cS}{{\cal S }}

\newcommand{\cV}{{\cal V }}

\newcommand{\cZ}{{\cal Z }}
\newcommand{\cO}{{\cal O }}
\newcommand{\cQ}{{{\cal Q }}}

\newcommand{\cJ}{{\cal J }}

\newcommand{\prt}{\partial}

\newcommand{\prtx}{{\partial_{(x)}}}

\newcommand{\yy}{{(y)}}

\newcommand{\xx}{{(x)}}

\newcommand{\xp}{{(x')}}

\newcommand{\zer}[1]{\stackrel{\circ}{#1}}

\newtheorem{Statement}{Statement}[section]

\newtheorem{Result}{Result}
\newtheorem{Ansatz}{Ansatz}

\begin{document}
\title{\bf{A model for massless higher spin field interacting  with a geometrical background.}}
\author{\bf{Giuseppe Bandelloni}\\
Physics Department of Genoa University,\\
and\\
Nuclear Physics National Institute, INFN,  Genoa Section\\
 via
Dodecaneso 33, I-16146  GENOVA Italy}
\maketitle
\footnote{keywords:{Quantum Field Theory
11.30.-j  Symmetry and conservation laws}\\
 pacs:{  11.10.-z;11.10.Cd;11.10.Ef;11.10.Gh}
 \\
email{ beppe@genova.infn.it}}
\begin{abstract}
{We  study a very general  four dimensional Field Theory model
 describing the dynamics of a massless higher spin $N$ symmetric tensor field particle interacting with a geometrical  background.This model is invariant under the action of an extended  linear diffeomorphism. We investigate  the consistency of the equations of motion,   and   the highest spin degrees of freedom are extracted by means of a set of  covariant constraints. Moreover the the highest spin equations of motions (and in general all the highest spin field  1-PI irreducible Green functions) are invariant under a chain of transformations induced by  a set of $N-2$ Ward operators, while  the auxiliary fields equations of motion spoil this symmetry.   The first steps to a quantum extension of the model are discussed on the basis of the Algebraic Field Theory.Technical aspects are reported in Appendices; in particular one of them  is devoted to illustrate the spin-$2$ case.}
\end{abstract}

\vskip 1.5cm
\section{Introduction-The spin decomposition for a Symmetric Tensor Field }
\lbl{Introduction}
\vskip 1.5cm
One of the challenging open questions in Field Theory concerns  the relativistic  Higher Spin gauge fields.

For a long time the attention  was focused on the construction of the Lagrangian of the Higher Spins within the Bargmann-Wigner \cite{Bargmann:1948ck} classification.
The fundamental papers of Fang and Frondal\cite{Fang:1978wz}\cite{Fronsdal:1978rb} opened new perspective for this problem, in the line  with the linearized gravity gauge field approach.

 Later on the String point of view was conceived as the more natural setting for this theme, since the strings vibrating spectrum  includes a tower  of fields with increasing spin.
This led to consider  mathematical progress
 in particular by the Russian school \cite{Fradkin:1987ks}\cite{Buchbinder:2002ry}\cite{Vasiliev:1990vu} (and many others...).

In this paper we present an approach which somehow  return to the   old fashioned style,
in the sense that relies more on the  the algebraic Field Theory, Symmetry  principles, and goes
beyond any linearization.

In four dimensions ,the right object to describe a spin $N$ particle is the symmetric tensor $\cA^{(\mu_{(1)},\cdots\mu_{(N)})}\xx$, however this field carries, besides  the the highest spin, many lower spins spaces which are  essential to insure  the relativistic covariance.
Therefore the first task is to  to identify  a procedure which extracts  the highest spin degrees of freedom and allows to study their dynamics without any contribution for the lowest spin sectors.

 To clarify this important point,  we repeat an argument of Reference \cite{Bandelloni:2011zz}, starting with a massive particle.

In the following table, for any tensor order,we report  the  $SO(3)$ subgroup decomposition:

\begin{center}
{\bf{Massive field intrinsic angular momentum decomposition}}
\begin{tabular}{|c|c|c|}
\hline Tensor Order & Dimensionality& $SO(3)$ Spin Decomposition \\ \hline
1&4& \{\textbf{1}\}+\{\textbf{0}\} \\ \hline
2&10& \{\textbf{2}\}+\{\textbf{1}\}+2\{\textbf{0}\} \\ \hline 3
&20&\{\textbf{3}\}+\{\textbf{2}\}+2\{\textbf{1}\}+2\{\textbf{0}\}\\
\hline 4 &35
&\{\textbf{4}\}+\{\textbf{3}\}+2\{\textbf{2}\}+2\{\textbf{1}\}+3\{\textbf{0}\}\\
\hline 5 &56&
\{\textbf{5}\}+\{\textbf{4}\}+2\{\textbf{3}\}+2\{\textbf{2}\}+3\{\textbf{1}\}+3\{\textbf{0}\}\\
\hline $\cdots$&$\cdots$&$\cdots$\\
\hline
N-1&$\binom{N+2}{N-1}$&\{\textbf{N-1}\}+\{\textbf{N-2}\}+2\{\textbf{N-3}\}+2\{\textbf{N-4}\}
+3\{\textbf{N-5}\}\\& &+3\{\textbf{N-6}\} +$\cdots$\\ \hline
N &$\binom{N+3}{N}$
&\{\textbf{N}\}+\{\textbf{N-1}\}+2\{\textbf{N-2}\}+2\{\textbf{N-3}\}
+3\{\textbf{N-4}\}\\& &+3\{\textbf{N-5}\} +$\cdots$
=$\sum_{j=0}^N
\Biggl[IntegerPart\biggl(\frac{N-j+2}{2}\Biggl)\Biggr]$\{\textbf{J}\}
\\
\hline
\end{tabular}

\lbl{tablemassive}

\end{center}
In order to isolate the highest spin $N$ term, we denote the whole field $\cA^{(\mu_{(1)},\cdots\mu_{(N)})}\xx$ by  the highest spin $N$ , $\cA^N\xx$
 its highest spin subspace as $\mathbf{S}^N \xx$. It is important to highlight in the general decomposition scheme the regularity of the growth of the multiplicity counting as we move towards the lowest momentum states.

To proceed, we have to remove the $\cA^{(N-1)}\xx$ tensor inside $\cA^N\xx$. At this stage we are left with an object which is composed with blocks (all with multiplicity 1) having the same parity of $\mathbf{S}^N \xx$. All these are removed by the term $({\cA}^{(N-2)}\xx-\cA^{(N-3)}\xx\Biggr)$ due the uniformity of the multiplicity decomposition content pointed out before.

So we get the important rule:

\begin{eqnarray}
\cA^{(N)}\xx&=&{\Biggl[\mathbf{S}^N\xx+
\cA^{(N-1)}\xx
+\Biggl({\cA}^{(N-2)}\xx -\cA^{(N-3)}\xx\Biggr)\Biggr]}\nn\\
\lbl{decompmassive}
\end{eqnarray}

Thus we see that the "extra" degrees of freedom carried by lower spin are carried by  tensors,
underlined by Weinberg\cite{Weinberg:1964cn} \cite{Weinberg:1964ev}\cite{Weinberg:1969di}.

This result allows to extract the highest spin by means of covariant supplementary conditions , as just known  (within different decompositions as the one in Equation \rf{decompmassive} and procedures) in References \cite{Fierz:1939ix}\cite{Fang:1978wz}\cite{Fronsdal:1978rb}.\footnote{For a more recent reviewing of the stats of the problem see Refs \cite{Sorokin:2004ie}\cite{Bouatta:2004kk}\cite{Bekaert:2005vh} and references quoted therein.}
 Notice that the decomposition in Equation \rf{decompmassive} holds for $N>3$, while for lower spin states, we point out that the related removal procedure could include pure scalar states: this is not our case, as we shall see   later in details.

Again Equation \rf{decompmassive} derives from the fact that the massive particle little group
 $O(d-1)$ representation admits, in d-dimensions, the decomposition:

\begin{eqnarray}
&&Dim^{Irreps}_{O(d-1)}=
\frac{2N+(d-3)}{N}\binom{(N-1)+(d-3)}{(N-1)} \nn\\
&=&\binom{N+(d-1)}{N}-\binom{(N-1)+(d-1)}{(N-1)}-\Biggl[\binom{(N-2)+(d-1)}{(N-2)}
-\binom{(N-3)+(d-1)}{(N-3)}\Biggr]\nn\\
&=&Dim^d_{(N)}-Dim^d_{(N-2)}-(Dim^d_{(N-1)}-Dim^d_{(N-3)})\nn\\
\lbl{dimlittlegroup}
\end{eqnarray}

 which suggests the solution for the massless particle case, where the little group is $E(2)$.  We can verify in Equation \rf{dimlittlegroup}, performing a virtual dimensional reduction, and going from $d=4$ to $d=3$,we get the right result for the physical states of the massless particle, with the same decomposition scheme as in Equation\rf{decompmassive}.

Indeed, removing from the symmetric
  $N$-th  tensor, its $N-1$-th  in $d$ dimensions, we get a new tensor whose  dimensionality is the one  of an $N$-th order symmetric tensor living in the $d-1$ dimensional space,
It is easy, in fact, to check:
\begin{eqnarray}
&&Dim^d_{(N)}-Dim^d_{(N-1)}=\binom{N+(d-1)}{(d-1)}-\binom{(N-1)+(d-1)}{(d-1)}=\binom{N+(d-2)}{(d-2)}
=Dim^{d-1}_{(N)}\nn\\
\lbl{dimensionsdown}
\end{eqnarray}
This signal that we need a $N-1$ order constraint for all  $\cA^{(\mu_1,\cdots,\mu_N)}\xx,\forall N$ field.
 The most suitable appears  to be the covariant Lorentz condition (transversality condition), so we impose:
\bigskip
\begin{eqnarray}
&&{\cV_1}^{(\mu_1,\cdots,\mu_{j})}\xx=\cD_{\mu_1} \cA^{(\mu_1,\cdots,\mu_j)}\xx=0 \quad j=1,\cdots, N
\lbl{Lorentz}
\end{eqnarray}

To confirm our argument,  we repeat the strategy given in the Table \rf{decompmassive}, performing now (for the massless case) the helicity decomposition of symmetric tensor which verify Equation \rf{Lorentz}: we obtain:

\begin{center}
{\bf{Massless field intrinsic angular momentum decomposition}}
\begin{tabular}{|c|c|c|}
\hline Tensor Order & Dimensionality& $E(2)$  Helicity Decomposition \\
\hline
1&3& \{\textbf{1}\}+\{\textbf{0}\} \\
\hline 2&6& \{\textbf{2}\}+\{\textbf{1}\}+2\{\textbf{0}\} \\ \hline 3
&10&\{\textbf{3}\}+\{\textbf{2}\}+2\{\textbf{1}\}+2\{\textbf{0}\}\\
\hline 4 &15
&\{\textbf{4}\}+\{\textbf{3}\}+2\{\textbf{2}\}+2\{\textbf{1}\}+3\{\textbf{0}\}\\
\hline
5 &21&
\{\textbf{5}\}+\{\textbf{4}\}+2\{\textbf{3}\}+2\{\textbf{2}\}+3\{\textbf{1}\}+3\{\textbf{0}\}\\
\hline $\cdots$&$\cdots$&$\cdots$\\
\hline
N-1&$ \binom{N+1}{N-1} $&\{\textbf{N-1}\}+\{\textbf{N-2}\}+2\{\textbf{N-3}\}+2\{\textbf{N-4}\}
+3\{\textbf{N-5}\}\\& &+3\{\textbf{N-6}\} +$\cdots$\\
 \hline  N
&$\binom{N+2}{N}$&\{\textbf{N}\}+\{\textbf{N-1}\}+2\{\textbf{N-2}\}+2\{\textbf{N-3}\}
+3\{\textbf{N-4}\}\\&&+3\{\textbf{N-5}\} +$\cdots$ =
$\sum_{j=0}^N
\Biggl[IntegerPart\biggl(\frac{N-j+2}{2}\Biggl)\Biggr]$\{\textbf{H}\}\\
\hline
\end{tabular}
\lbl{tablemassless}
\end{center}
so we conclude that also in the massless case (which obey the condition \rf{Lorentz}),{ \it{we have  the same decomposition mechanism obtained Equation \rf{decompmassive}, as predicted also by Equation \rf{dimlittlegroup}}}

In order to implement the procedure to isolate the highest spin state we introduce auxiliary fields
 $\cA^{(\mu_1,\cdots,\mu_j)}\xx$($j<N$) and associate them to the  $\cA^{(\mu_1,\cdots,\mu_N)}\xx$ subtensors   by means suitable constraint conditions.

 The choice of them is not unique; adopting here a simplicity criterium, we impose:

\begin{eqnarray}
&&{\cV_2}^{(\mu_1,\cdots,\mu_{N-1})}\xx= a_{\mu_N}\xx\cA^{(\mu_1,\cdots,\mu_N)}\xx-
\cA^{(\mu_1,\cdots,\mu_{N-1})}\xx=0\nn\\
\lbl{sectionmassive1}\\
&& {\cV_3}^{(\mu_1,\cdots,\mu_{N-2})}\xx=a_{(\mu_N,\mu_{N-1})}\xx\cA^{(\mu_1,\cdots,\mu_{N})}\xx
-\cA^{(\mu_1,\cdots,\mu_{N-2})}\xx=0\lbl{sectionmassive2}\\
&&{\cV_4}^{(\mu_1,\cdots,\mu_{N-3})}\xx = b_{(\mu_{N-2})}\xx\cA^{(\mu_1,\cdots,\mu_{N-2})}\xx-\cA^{(\mu_1,\cdots,\mu_{N-3})}\xx=0
\lbl{sectionmassive3}
\end{eqnarray}

with local parameters $a_{\mu_N}\xx$,$a_{(\mu_N,\mu_{N-1})}\xx$,$b_{(\mu_{N-2})}\xx$.

In the massless case, we must add the condition \rf{Lorentz} to the  the previous conditions \rf{sectionmassive1},\rf{sectionmassive2},\rf{sectionmassive3}.

 Indeed Equation\rf{Lorentz} leave us with  $\binom{N+2}{2}$ free degrees of freedom. On the other hand Equations\rf{sectionmassive1} imply $\binom{N+1}{2}$ constraints,moreover equations \rf{sectionmassive2} fix $\binom{N}{2}$ conditions, to which we must remove $\binom{N-1}{2}$ degrees of freedom from equations \rf{sectionmassive3}.
At this point we are left with only {\it two} residual degrees of freedom which surely describes the top spin
helicity since the removal procedure enlist only states of lower angular momentum.

At a final point we want to point out that,for lowest spins,where the fields chain is not complete (s=1,2,3),
no scalar field is needed for this purpose.

1)$\bullet$  $s=1$: the tree degrees of freedom of the $\cA^{(\mu)}\xx$ field  are reduced to two by the condition $a_{(\mu)}\xx\cA^{(\mu)}\xx=0$

2) $\bullet$  $s=2$: the six degrees of freedom of the $\cA^{(\mu,\nu)}\xx$ field are first reduced to $3$ by the conditions $a_{(\mu)}\xx\cA^{(\mu,\nu)}\xx=\cA^{(\nu)}\xx$, and finally to two using the condition $a_{(\mu,\nu)}\xx\cA^{(\mu,\nu)}\xx=0$

3) $\bullet$ $s=3$: the $10$ free degrees of freedom of  $\cA^{(\mu,\nu,\rho)}\xx$ field are first reduced to $4$ using the six equations\\ $a_{(\mu)}\xx\cA^{(\mu,\nu,\rho)}\xx=\cA^{(\nu,\rho)}\xx$ and then to $2$ using the two constraints $a_{(\mu,\nu)}\xx\cA^{(\mu,\nu,\rho)}\xx=\cA^{(\rho)}\xx$, since the vector $\cA^{(\rho)}\xx$ field must be constrained using the incomplete \rf{sectionmassive3} equation $b_{(\rho)}\xx\cA^{(\rho)}\xx=0$.

Ascending the spin content the conditions \rf{sectionmassive1},\rf{sectionmassive2},\rf{sectionmassive3} are fully restored.
\vskip 0.5 cm

We are now ready to rephrase the above formulation for a Lagrangian field theory with an Action functional $S$.
The field $\cA^{(\mu_1,\cdots,\mu_N)}\xx$ obeys the evolve the equations of motion:
\begin{eqnarray}
\frac{\delta S}{\delta \cA^{(\mu_1,\cdots,\mu_N)}\xx}=0
\lbl{topspinmotion}
\end{eqnarray}
which describe the evolution of all the spin content contained in $\cA^{(\mu_1,\cdots,\mu_N)}\xx$.

In order to extract only the highest spin we introduce the functional counterparts of Equations
\rf{sectionmassive1},\rf{sectionmassive2},\rf{sectionmassive3} i.e:

\begin{eqnarray}
\Biggl[{{a}^{-1}}^{(\rho_N)}\xx\frac{\delta }{\delta \cA^{(\mu_1,\cdots,\mu_N)}\xx}-\frac{\delta }{\delta \cA^{(\mu_1,\cdots,\mu_{N-1})}\xx}\Biggr]S=0\nn\\
\lbl{N-1}
\end{eqnarray}
{\footnote{We have introduced  the inverse ${{a^{}}^{-1}}^{(\nu_{1},\cdots,\nu_{j})}\xx$ defined as:

$${{a^{}}^{-1}}^{(\nu_{1},\cdots,\nu_{j})}\xx = \frac{1}{{|a|}^2}a^{}_{(\mu_1,\cdots,\mu_{j})}\xx\delta^{
((\nu_{1},\cdots,\nu_{j}),(\mu_{1},\cdots,\mu_{j}))}$$

and similarly for $b$}
}

we remove the $N-1$ tensorial spin contribution,and
we are left with  $\binom{N+2}{N}$ unconstrained equations.
The next step is worked out putting:
\begin{eqnarray}
\Biggl[{{a}^{-1}}^{(\rho_N,\rho_{N-1})}\xx\frac{\delta }{\delta \cA^{(\mu_1,\cdots,\mu_N)}\xx}-\frac{\delta }{\delta \cA^{(\mu_1,\cdots,\mu_{N-2})}\xx}\Biggr]S=0\nn\\
\lbl{N-2}
\end{eqnarray}

where the  $\binom{N+1}{N-2}$ degrees of freedom of the term:
\begin{eqnarray}
\frac{\delta S}{\delta \cA^{(\mu_1,\cdots,\mu_{N-2})}\xx}
\end{eqnarray}
is processed with the subsidiary constraints:
\begin{eqnarray}
\Biggl[{{b}^{-1}}^{(\rho_{N-2})}\xx\frac{\delta }{\delta \cA^{(\mu_1,\cdots,\mu_{N-2})}\xx}-\frac{\delta }{\delta \cA^{(\mu_1,\cdots,\mu_{N-3})}\xx}\Biggr]S=0\nn\\
\lbl{N-3}
\end{eqnarray}
and then  are reduced to  $\binom{N}{N-2}$.

If so, the constraints imposed by equations \rf{N-2} reduces the $\binom{N+2}{N}$ degrees of freedom of $\frac{\delta S}{\delta \cA^{(\mu_1,\cdots,\mu_{N-1})}\xx}$, by means the residual states survived to the action of Equation \rf{N-3}.
So we are left with  $\binom{N+2}{N}-\binom{N}{N-2}=2N+1$  degrees of freedom of highest spins components of $\frac{\delta S}{\delta \cA^{(\mu_1,\cdots,\mu_{N})}\xx}$.

\bigskip
In the massless case, we  also have to impose the covariant transversality condition:

\begin{eqnarray}
{\cD_{(\mu_j)}}\frac{\delta S}{\delta \cA^{(\mu_1,\cdots,\mu_j)}\xx}=0\quad j=1,\cdots,N\nn\\
\lbl{Lorentzconditionmotion}
\end{eqnarray}

and then repeat
 the steps \rf{N-1},\rf{N-2},\rf{N-3}, to  isolate the $2$ unconstrained helicity for the highest spin field $\frac{\delta S}{\delta \cA^{(\mu_1,\cdots,\mu_{N})}\xx}$.

\vskip 1.0cm

The above procedure, which is clearly covariant, might not be valid (a priory) in all reference frame since takes its origin from a "non relativistic " analysis.

In order to insure its validity we also impose the additional symmetry of reparametrization invariance.

This symmetry is carried by a linear extension\cite{Fang:1978rc}  of the diffeomorphism algebra.

We implement this program by the following Ansatz

\begin{Ansatz}
\lbl{Ansatz}
The local linear  transformation  we adopt here operates  on the multiplet  $\cA^{(\mu_1,\cdots,\mu_j)}\xx$ $j=1,\cdots,N$ such that:

\begin{eqnarray}
 &&\cA^{(\mu_1,\cdots,\mu_j)}\xx \lra{\cA^\prime}^{(\mu_1,\cdots,\mu_j)}\xx=\cA^{(\mu_1,\cdots,\mu_j)}\xx\nn\\
  &&+\int d^4 x'\sum_{j,i=1,\cdots, N; i\leq j}\cF^{(\mu_1,\cdots,\mu_j)}_{(\nu_1,\cdots\nu_i)}(x|x')\cA^{(\nu_1,\cdots\nu_j)}\xp\nn\\
 \lbl{gaugemapping}
 \end{eqnarray}

 Note that, due to the tensorial character of $\cF^{(\mu_1,\cdots,\mu_j)}_{(\nu_1,\cdots\nu_i)}(x|x') $ , the transformed fields depend on the untransformed ones  with equal or lower degree, which correspond to the physical requirement that the highest spin components should not appear in the transformed lower order transformed fields.

 \end{Ansatz}

Two points are now to be discussed: first the compatibility of Equations
 \rf{Lorentz},\rf{sectionmassive1},\rf{sectionmassive2},\rf{sectionmassive3}
 with the symmetry of Ansatz \rf{Ansatz}, and in particular the independence of the cohomology space from the local parameters $a_{(\mu)}\xx,a_{(\mu,\nu)}\xx,b_{(\mu)}\xx$.
Second, what happens to the physical degrees of freedom after the introduction of this symmetry,
 and, then, the presence of equivalence classes among the states.

 {{Recall that we are dealing with tensorial fields $ \cA^{(\mu_1,\cdots,\mu_j)}\xx; j=1,\cdots N$ and no scalar field is needed.

 Looking at the massless case, the constraints in Equations \rf{Lorentz},\rf{sectionmassive1},\rf{sectionmassive2},\rf{sectionmassive3}, lead to the counting:
  \begin{eqnarray}
\textrm{Total degrees of freedom} - \textrm{Symmetry parameters}
-\textrm{ Constraints }= \textrm{Physical degrees of
freedom}
\lbl{fundamental}
\end{eqnarray}

therefore we conclude that to be consistent with the previous analysis, the number of the symmetry parameters must be equal to the one of the auxiliary fields.

Hence, if we wish to include reparametrization invariance, we forced to have a symmetry algebra described by tensorial parameters with all orders from $1$ to $N-1$.
}}

We shall see later that this provides a simple link between this symmetry
 and the one of Fang and Fronsdal.

 Finally we should also analyze if the functional constraints constraints in Equations \rf{sectionmassive1},\rf{sectionmassive2},\rf{sectionmassive3} are compatible with the symmetry transformation.

This will be done by standard gauge gauge models techniques, since the constraints freeze the lower spin sectors in much the  same way as the gauge symmetry separates the unphysical degrees of freedom from the physical ones.

\vskip 1.0 cm

The paper is organized as follows:

\vskip 0.5 cm
$\bullet$ In Section \rf{symmetry} we build the B.R.S. algebra connected with the Ansatz \rf{Ansatz}. The link with the Fang-Fronsdal \cite{Fronsdal:1978rb} symmetry is easily derived.

\vskip 0.5 cm

$\bullet$ In Section \rf{model} we build a general massless Field Theory Lagrangian model, whose principal ingredient is the multiplet $\cA^{(\mu_1,\cdots,\mu_j)}\xx; j=1,\cdots, N$ interacting with an external geometrical background. The constraints found in in this Section, and the ones necessary to select the highest spin interactions, are included via a linear  fixing procedure process. The model is algebraically  studied only at the Classical level.The top spin field equation equation of motion (and in general all the 1-PI irreducible top spin Green functions) are invariant under a family of Ward operators.

\vskip 0.5 cm

$\bullet$ In Section \rf{conclusions} we briefly sum up our conclusions, in view of the future objectives we have to achieve.
\vskip 0.5 cm
$\bullet$  The Appendices are devoted to technicalities: In  Appendix \rf{extendederivative} we rephrase the symmetry of Section \rf{symmetry},which is particularly suitable for cohomological aspects. In Section\rf{motionconstraints} we derive the constraints fulfilled by the equations of motion, to have a well defined dynamics for the highest spin field. In the sub-appendix \rf{spin2} we completely analyze the equations of motion of the spin-$2$ case, and the extraction of the highest spin component, by means of an interesting algebraic condition.

\section{Symmetry Tranformations}
\lbl{symmetry}

In this Section we investigate the  of the assumptions in the Ansatz \rf{Ansatz}: we derive here the algebra of the transformations which will identify the physics.

The most natural solution, which links with the current Literature, is a tensorial symmetry which includes the Fang-Fronsdal \cite{Fang:1978wz} one.
This one comes from the extension of the linearized reparametrization algebra.

We recall that the coordinates reparametrization transformations of $\cA^{(\mu_1,\cdots,\mu_N)}\xx$, are defined,  a la B.R.S \cite{Becchi:1975nq}, as:
\begin{eqnarray}
\delta\cA^{(\mu_1,\cdots,\mu_N)}\xx&=&\int d^4 x' d^4 x"\sum_{i=1}^N f^{\mu_i}_{[\nu,\sigma]}(x,[x',x"])\cA^{(\mu_1,\cdots_{\widehat{\mu_i}}\sigma,\cdots,\mu_N)}
(x")\ccC^\nu(x')\nn\\
\lbl{diff}
\end{eqnarray}
with structure functions\cite{DeWitt:1965jb}:
\begin{eqnarray}
f^\mu_{[\nu,\sigma]}(x,[x',x"])&=&\delta^\mu_\sigma \biggl(\prtx_\nu\delta(x-x")\biggr)\delta(x-x')-\delta^\mu_\nu\biggl(
\prtx_\sigma\delta(x-x')\biggr)\delta(x-x")\nn\\
\lbl{diffstructurefunctions}
\end{eqnarray}
so that:

\begin{eqnarray}
\delta\cA^{(\mu_1,\cdots,\mu_N)}\xx&=&\ccC^\lambda\xx\prt_\lambda \cA^{(\mu_1,\cdots,\mu_N)}\xx-\sum_{i=1}^N\prt_\lambda\ccC^{(\mu_i)}\xx\cA^{(\mu_1,\cdots_{\widehat{\mu_i}}\lambda,\cdots,\mu_N)}\xx\nn\\
\end{eqnarray}

This extension was studied by the present author \cite{Bandelloni:2011zz}within a symplectic scenario, however it can be applied to a whatever geometrical environment.  Many mathematical details  are discussed in the previous reference.

This approach gives a  linear tensorial extensions with the structure constants:

\begin{eqnarray}
f^{(\mu_1,\cdots,\mu_j)}_{[(\nu_1,\cdots,\nu_{j-k+1}),(\sigma_1\cdots,\sigma_{k})]}(x,[x',x"])&=&
\sum_{l=1}^{j-k+1}\delta^{(\mu_1,\cdots,\mu_j)}_{((\nu_1,\cdots
_{\widehat{\nu_l}}
\cdots,\nu_{j-k+1}),(\sigma_1\cdots,\sigma_{k}))}
\biggl(\prtx_{\nu_l}\delta(x-x")\biggr)\delta(x-x')\nn\\
&-&\sum_{l=1}
^{k}
\delta^{(\mu_1,\cdots,\mu_j)}_{((\nu_1,\cdots\nu_{j-k+1}),
(\sigma_1\cdots
_{\widehat{{\sigma}_{l}}}
\cdots,\sigma_{k}))}
\biggl(
\prtx_{\sigma_l}\delta(x-x')\biggr)\delta(x-x")\nn\\
\lbl{structurefunctions}
\end{eqnarray}

This algebra  implies an enlargement of the number of fields.We have previously seen that, in the case with $N>4$, and for $N=2$, the lowest spin states removal procedure is performed by the aid of auxiliary tensor fields.
 In this case we introduce  the whole tower of symmetric tensor fields$\cA^{(\mu_1,\cdots,\mu_j)}\xx$ with  rank $j$ spanning from one to $N$,So the element  of the algebra must be represented by symmetric tensor parameters  $\ccC^{(\nu_1,\cdots,\nu_{j})}(x')$ (which will be embedded in a B.R.S. framework as ghosts fields) whose rank, if we consider the total number of tensor  fields,algebra parameters and constraints, must  span from one to $N-1$.In the other cases ($N=3,4$)we have to introduce auxiliary scalar states, so the naive degrees of freedom counting procedure requires   the addition of scalar algebra parameters, or, if we want to keep the previous algebra, we have to introduce some ad-hoc conditions.

So for a symmetric tensor field we perform the infinitesimal transformation:
\begin{eqnarray}
\delta\cA^{(\mu_1,\cdots,\mu_j)}\xx&=&\int d^4 x'  d^4 x" \sum_{k=1}^j
f^{(\mu_1,\cdots,\mu_j)}_{[(\nu_1,\cdots,\nu_{j-k+1}),(\sigma_1\cdots,\sigma_{k})]}(x,[x',x"])\cA^
{(\sigma_1,\cdots,\sigma_k)}(x")\ccC^{(\nu_1,\cdots,\nu_{j-k+1})}(x')\nn\\
\end{eqnarray}

which gives:

\begin{eqnarray}
\delta\cA^{(\mu_1,\cdots,\mu_j)}\xx&=&\delta^{(\mu_1,\cdots,\mu_j)}_{(\rho_1,\cdots,\rho_j)}
\Biggl[\sum_{i=0}^{j-1}
\Biggl(\ccC^{(\lambda,\rho_{1},\cdots,\rho_{i})}\xx\prt_{(\lambda)}
\cA^{(\rho_{i+1},\cdots,\rho_j)}\xx
-\cA^{(\lambda,\rho_{1},\cdots,\rho_{i})}
\xx\prt_{(\lambda)}
\ccC^{(\rho_{i+1},\cdots,\rho_j)}\xx\Biggr)\Biggr]\nn\\
j&=&1,\cdots,N\nn\\
\lbl{SA}
\end{eqnarray}
in  particular  the top field transforms as:

\begin{eqnarray}
&&\cS \cA^{(\mu_{(1)}),\cdots,\mu_{(N)})}\xx=
\ccC^{(\lambda)}\xx\prt_{(\lambda)}\cA^{(\mu_{(1)}),\cdots,\mu_{(N)})}\xx
+\delta^{(\mu_{(1)}),\cdots,\mu_{(N)})}_{{(\rho_{(1)}),\cdots,\rho_{(N)})}}
\sum_{j=1}^{N-2}\ccC^{(\nu,\rho_1,\cdots,\rho_j)}\xx
\prt_{(\nu)}\cA^{(\rho_{j+1},\cdots,\rho_N)}\xx\nn\\
&-&
\delta^{(\mu_{(1)}),\cdots,\mu_{(N)})}_{{(\rho_{(1)}),\cdots,\rho_{(N)})}}
\sum_{j=1}^{N-2}\cA^{(\nu,\rho_1,\cdots,\rho_j)}\xx
\prt_{(\nu)}\ccC^{(\rho_{j+1},\cdots,\rho_N)}\xx\nn\\
\end{eqnarray}
Note that, on the basis of the exclusion of the $\ccC^{(\mu_{(1)}),\cdots,\mu_{(N)})}\xx$ ghosts, seen before,  the B.R.S. variation of  the top field
 $\cA^{(\mu_{(1)}),\cdots,\mu_{(N)})}\xx$
 does not  include in the list the first order one
 $\cA^{(\mu)}\xx$.

 The chain algebra ends with:
\begin{eqnarray}
&&\cS \cA^{(\mu,\nu,\rho)}\xx=\ccC^{(\lambda)}\xx \prt_{(\lambda)}
\cA^{(\mu,\nu,\rho)}\xx-\prt_{(\lambda)}
\ccC^{(\mu)}\xx\cA^{(\lambda,\nu,\rho)}\xx-\prt_{(\lambda)}\ccC^{(\nu)}\xx
\cA^{(\mu,\lambda,\rho)}\xx\nn\\&&-\prt_{(\lambda)}\ccC^{(\rho)}\xx\cA^{(\mu,\nu,\lambda)}\xx
-\prt_{(\lambda)}\ccC^{(\mu,\nu)}\xx\cA^{(\lambda,\rho)}\xx-\prt_{(\lambda)}
\ccC^{(\mu,\rho)}\xx\cA^{(\lambda,\nu)}\xx-\prt_{(\lambda)}
\ccC^{(\rho,\nu)}\xx\cA^{(\lambda,\mu)}\xx\nn\\
&&+\prt_{(\lambda)}\cA^{(\mu,\nu)}\xx\ccC^{(\lambda,\rho)}\xx+
\prt_{(\lambda)}\cA^{(\mu,\rho)}\xx\ccC^{(\lambda,\nu)}\xx+\prt_{(\lambda)}
\cA^{(\rho,\nu)}\xx\ccC^{(\lambda,\mu)}\xx-\cA^{(\lambda)}\xx\prt_{(\lambda)}
\ccC^{(\mu,\nu,\rho)}\xx\nn\\
&&+\prt_{(\lambda)}\cA^{(\mu)}
\xx\ccC^{(\lambda,\nu,\rho)}\xx+\prt_{(\lambda)}\cA^{(\nu)}\xx
\ccC^{(\mu,\lambda,\rho)}\xx+\prt_{(\lambda)}\cA^{(\rho)}\xx
\ccC^{(\mu,\nu,\lambda)}\xx\nn\\
&&\nn\\
&&\cS \cA^{(\mu,\nu)}\xx=\ccC^{(\lambda)}\xx \prt_{(\lambda)}
\cA^{(\mu,\nu)}\xx-\cA^{(\lambda,\nu)} \xx \prt_{(\lambda)}
\ccC^{(\mu)}\xx
-\cA^{(\mu,\lambda)}\xx \prt_{(\lambda)}\ccC^{(\nu)}\xx\nn\\
&&-\cA^{(\lambda)}\xx
\prt_{(\lambda)}\ccC^{(\mu,\nu)}\xx+\ccC^{(\mu,\lambda)}\xx
\prt_{(\lambda)}\cA^{(\nu)}\xx+\ccC^{(\nu,\lambda)}\xx\prt_{(\lambda)}\cA^{(\mu)}\xx
\nn\\
&&\nn\\
&&\cS \cA^{(\mu)}\xx=\ccC^{(\lambda)}\xx\prt_{(\lambda)}
\ccA^{(\mu)}\xx -\cA^{(\lambda)}
\xx \prt_{(\lambda)}\ccC^{(\mu)}\xx\nn\\
\nn\\
\end{eqnarray}

We underline that the top field ($\cA^{(\rho_1,\cdots,\rho_j)}\xx;j=N$ )terms is absent in the B.R.S transformations of the fields with $j<N$

If we adopt a B.R.S.\cite{Becchi:1975nq} framework, the $\ccC^{(\mu_1,\cdots,\mu_j)}\xx$ ghost parameters have a  variation:
 \begin{eqnarray}
\cS\ccC^{(\mu_1,\cdots,\mu_j)}\xx
&=& -\frac{1}{2}\int d^4 x'  d^4 x" \sum_{l=1}^j f^{(\mu_1,\cdots,\mu_j)}_{[(\nu_1,\cdots,\nu_{l}),(\sigma_1\cdots,\sigma_{j-l+1})]}(x,[x',x"])\ccC^
{(\sigma_1,\cdots,\sigma_{j-l+1})}(x")\ccC^{(\nu_1,\cdots,\nu_{l})}(x')\nn\\
&=&\delta^{(\mu_1,\cdots,\mu_j)}_{(\rho_1,\cdots,\rho_j)}
\Biggl[\sum_{i=0}^{j-1}
\ccC^{(\lambda,\rho_{1},\cdots,\rho_{i})}\xx\prt_{(\lambda)}
\ccC^{(\rho_{i+1},\cdots,\rho_j)}\xx\Biggr]\nn\\
j&=&1,\cdots,N-1\nn\\
\lbl{SC}
\end{eqnarray}

It is clearly visible that the algebra is arranged into growing sectors which stop at the order of the highest spin field.

An (alternative) description of this algebra, which enhances its value, can be done introducing, in a Fock space, the $j-th$  generalized tensor derivative operator:

\begin{eqnarray}
\prt_{(\mu_1\cdots,\mu_j)}\equiv \Biggl\{\frac{\prt}{\prt
\ccC^{{(\mu_1\cdots,\mu_j)}}\xx},\delta \Biggr\} \lbl{derextended}
\end{eqnarray}
An Appendix \rf{extendederivative} is devoted to this notation,which enormously simplify the algebraic treatment.

For example  Equations \rf{SC} and \rf{SA} transformations can be rewritten in this scheme as:

\begin{eqnarray}
\cS \ccC^{(\rho_1,\cdots,\rho_k)}\xx
=\sum_{l=1}^k\ccC^{(\mu_1,\cdots,\mu_l)}\xx\prt_{(\mu_1,\cdots,\mu_l)}
\ccC^{(\rho_1,\cdots,\rho_k))}\xx\nn\\
\lbl{affinity1}
\end{eqnarray}

\begin{eqnarray}
\cS \ccA^{(\rho_1,\cdots,\rho_k)}\xx =
\sum_{l=1}^N\ccC^{(\mu_1,\cdots,\mu_l)}\xx\prt_{(\mu_1,\cdots,\mu_l)}
\ccA^{(\rho_1,\cdots,\rho_k)}\xx-\ccA^{(\mu_1,\cdots,\mu_l)}\xx
\prt_{(\mu_1,\cdots,\mu_l)}\ccC^{(\rho_1,\cdots,\rho_k)}\xx\nn\\
\lbl{affinity2}
\end{eqnarray}

so the link with the reparametrization transformations is clear.
The use of macro-indices   reveals the underlying skeleton diffeomorphical nature of this symmetry.

Within this notation the cohomology calculations are more direct.
First of all, by means the use of the so called "descent equations"\cite{Stora:1983ct}, it is possible to relate the cohomology space on functionals to the one on functions.  Remark that, after the introduction of the extended derivatives technique: the differential to be used to graduate the  space of forms on local functions, has to be defined  as ${\bf d}\equiv \biggl\{\frac{\prt}{\prt \ccC^\mu\xx, \delta_{B.R.S.}}\biggr\}dx^\mu$: its extension to higher ghosts derivatives would imply the use of symmetric measures which do not fit the wedge structure of the one of the relativistic Action.
Then using the well-known spectral sequences techniques already \cite{Dixon:1979bs}, we can retrace the steps of References \cite{Bonora:1984ic} \cite{Bandelloni:1988ws} (and the complete calculation can be found in Ref \cite{Bandelloni:2011zz}), where the anomaly takes origin  from monomials build with the traces  of products of first order (generalized) derivatives of ghost fields, and in the four dimensional case we have the object:
\begin{eqnarray}
\sum_{j_i}\prt_{(\mu^1_1,\cdots\mu^1_{j_1})}\ccC^{(\mu^2_1,\cdots\mu^2_{j_2})}\xx
\prt_{(\mu^2_1,\cdots\mu^2_{j_2})}\ccC^{(\mu^3_1,\cdots\mu^3_{j_3})}
\prt_{(\mu^3_1,\cdots\mu^3_{j_3})}\ccC^{(\mu^4_1,\cdots\mu^4_{j_4})}
\prt_{(\mu^4_1,\cdots\mu^4_{j_4})}\ccC^{(\mu^5_1,\cdots\mu^5_{j_5})}
\prt_{{(\mu^5_1,\cdots\mu^5_j{j_5})}}\ccC^{(\mu^1_1,\cdots\mu^1_{j_1})}\xx\nn\\
\end{eqnarray}
But, if we consider these trace term, and if we impose  the  indices coincidence of the first left derivative and the last right ghost field , the previously mentioned  downgrading index process can be avoided only if all the orders of derivatives and ghost fields is equal to one. So anomalies in ghosts with order greater than one are absent.
Then, ascending the "descent equations" we get as only cohomological obstruction on functionals the well known gravitational anomaly \cite{Alvarez-Gaume:1983ig}.
This result is helpful for the symmetry quantum extension.

Finally,the   contact point of our algebra with the Fang-Fronsdal \cite{Fronsdal:1978rb} symmetry, which,at the present, enjoys great fortune   in the current Literature  must be recovered in a simple way: if we linearize:
\begin{eqnarray}
\cA^{(\mu_1,\cdots,\mu_j)}\xx&=&\zer{\cA^{(\mu_1,\cdots,\mu_j)}}+\tilde{\cA}^{(\mu_1,\cdots,\mu_j)}\xx \nn\\
\lbl{fangfronsdalimit}
\end{eqnarray}
and the lowest order term $\zer{\cA^{(\mu_1,\cdots,\mu_j)}}$ is zero unless the rank two term , which is fixed $\zer{\cA^{(\mu_1,\mu_2)}}=\delta^{(\mu_1,\mu_2)}$,then
 we immediately derive, at the zeroth order, the Fang-Fronsdal algebra:

\begin{eqnarray}
\cS\cA^{(\mu_1,\cdots,\mu_j)}\xx&=&-\delta^{(\mu_1,\cdots,\mu_j)}_{(\nu_1,\cdots,\nu_j)}
\prt^{\nu_1}\ccC^{(\nu_2,\cdots,\nu_j)}\xx\nn\\
\lbl{fangfronsdalimit1}
\end{eqnarray}
 where the nilpotency condition trivializes the ghost algebra to be
  abelian:
\begin{eqnarray}
\cS{\ccC}^{(\rho_1\cdots,\rho_j)}\xx&=&0\nn\\
\lbl{fangfronsdalimit2}
\end{eqnarray}

The geometrical  scenarios of this algebra can be found in several References, for example Refs:  \cite{Francia:2002aa}\cite{Francia:2002pt}.

Furthermore the use of the structure constant in  Equation \rf{structurefunctions} for the transformations of low indices objects $ \cO_{(\mu_{1}\cdots \mu_{i})}\xx$ as:
\begin{eqnarray}
&&\cS \cO_{(\mu_{1}\cdots \mu_{i})}\xx =\sum_{l=1}^{N} \int
f^{(\rho_{1}\cdots \rho_{m})}_{[(\mu_{1}\cdots
\mu_{i}),(\nu_{1}\cdots
  \nu_{l})]}([x',x''],x) \cO_{(\rho_{1}\cdots \rho_{m})}(x'')\ccC^{(\nu_{1}\cdots \nu_{l})}
\xp d x' d x'' \nn\\
\end{eqnarray}
  shows that they are tensor densities.

   This fact will be used  for the Lagrangian construction.

 The usual covariant derivative of the tensor fields is defined as:

\begin{eqnarray}
\cD_\nu\cA^{(\mu_1,\cdots,\mu_j)}\xx=\prt_\nu\cA^{(\mu_1,\cdots,\mu_j)}\xx+
\sum_{m=1}^j\delta^{(\mu_1,\cdots,\mu_j)}_{(\rho_1,\cdots,\rho_j)}
\Gamma_{(\nu,\lambda)}^{(\rho_1,\cdots,\rho_m)}\xx\cA^{(\lambda,\rho_{m+1},\cdots,\rho_j)}\nn\\
\end{eqnarray}

We remark the presence of higher order connections and curvatures, already well-known in higher spin models.

The geometry we use here is studied in Ref \cite{Bandelloni:2011zz} with the introduction  of higher order metric tensors ${g}^{((\rho_1,\cdots\rho_j,\sigma_1,\cdots,\sigma_k))}\xx$; $j,k=1,\cdots,N$,
 which ,for $i,j>1$,  are not symmetric a priori (remember; the volume is a triple product) and verify the metricity conditions  $\cD_\mu g^{((\rho_1,\cdots\rho_j,\sigma_1,\cdots,\sigma_k))}\xx=0$
 The origin of this approach takes origin from Riemann.
{\footnote{ Georg Friedrich Bernhard Riemann in its work "Über die Hypothesen, welche der Geometrie zu Grunde liegen" ( ("On the hypotheses which underlie geometry")1854) said: A similar path to the same goal could also be taken in those manifolds in which the line element is expressed in a less simple way, e.g., by a fourth root of a differential expression of the fourth degree…
 . }}.

 Their B.R.S variations intrduces into the game the lower order tensors,much in the same way as the n-orders volumes infinitesimal deformations depend on  the lower  order volumes contained inside.
 The geometry is  richer and complex than the usual description one, and it is not treated here.{\footnote{For a fast reading we address the reader to "http://mathpages.com/rr/s9-03/9-03.htm"}} An  investigation on this subject  is under way.

Notice that this aspect is very important in relation to many physical themes such as or the Velo Zwanziger problem \cite{Velo:1970ur}the Weinberg-Witten theorem \cite{Weinberg:1980kq}and many others on the higher spin fields interactions \cite{Buchdahl:1958xv}\cite{Bekaert:2002uh}. One can immediately verify that our approach immediately justify the helicity flip of an Higher Spin massless particle (spin grater than two and  without any multiple interaction process ) interacting with the gravitational background.

The higher order metric description of these connections is
defined  recursively from the metricity conditions ( in Ref \cite{Bandelloni:2011zz})
as:

\begin{eqnarray}
\Gamma^{(\lambda_1,\cdots,\lambda_k)}_{(\rho,\eta)}\xx&=&\frac{1}{2}\Biggl[{\biggl
(g\biggr)}^{-1}_{(\mu,\eta)}\xx\Biggl(\prt_{(\rho)}
g^{((\lambda_1,\cdots\lambda_k,\mu))}\xx-
\Gamma^{(\mu)}_{(\rho,\sigma)}\xx
g^{((\lambda_1,\cdots\lambda_k,\sigma))}\xx\nn\\&-&
\delta^{(\lambda_1,\cdots,\lambda_k)}_{(\sigma_1,\cdots,\sigma_k)}
\sum_{i=1}^{k-1}\Gamma^{(\sigma_1,\cdots,\sigma_i)}_{(\rho,\nu)}\xx
g^{((\nu,\sigma_{i+1},\cdots,\sigma_k,\mu))}\xx
\Biggr)+\rho\longleftrightarrow\eta\Biggr] \lbl{Gamma}
\end{eqnarray}
( they are symmetric in the lower indices)
 whose  B.R.S.  variations are:
\begin{eqnarray}
&&\cS\Gamma_{(\sigma,\lambda)}^{(\mu_1,\cdots,\mu_j)}\xx=
\delta^{(\mu_1,\cdots,\mu_j)}_{(\rho_1,\cdots,\rho_j)}
\Biggl[\sum_{l=1}^j \sum_{r=0}^{j-l} \Biggl(
\ccC^{(\eta,\rho_{(l+1)}),\cdots,\rho_{(j)})}\xx\prt_{(\eta)}\Gamma_{(\sigma,\lambda)}^{(\rho_1,\cdots,\rho_l)}\xx
\nn\\&&+\prt_{(\sigma)}\ccC^{(\eta,\rho_{(l+1)}),\cdots,\rho_{(j)})}\xx
\Gamma_{(\eta,\lambda)}^{(\rho_1,\cdots,\rho_l)}\xx +
\prt_{(\lambda)}\ccC^{(\eta,\rho_{(l+1)}),\cdots,\rho_{(j)})}\xx\Gamma_{(\sigma,\eta)}^{(\rho_1,\cdots,\rho_l)}\xx
\nn\\&-&\prt_{(\eta)}\ccC^{(\rho_1,\cdots,\rho_l)}\xx\Gamma^{(\eta,\rho_{(l+1)}),\cdots,\rho_{(j)})}_{(\eta,\lambda)}\xx
\Biggr)\Biggr]
+\prt_{(\sigma)}\prt_{(\lambda)}\ccC^{(\mu_1,\cdots,\mu_j)}\xx
\nn\\
\lbl{sconnections}
\end{eqnarray}

With these definitions it easy to verify that the previous transformations in Equations \rf{SA}, \rf{SC} are unchanged after the trivial substitution of the ordinary derivative operators with the covariant ones.

The  higher order curvatures are canonically defined as:

\begin{eqnarray}
&&\cR_{\{[\lambda,\sigma],\eta\}}^{(\mu_1,\cdots,\mu_i)}\xx=\frac{\prt}{\prt
x^{(\lambda)}} \Gamma_{(\sigma,\eta)}^{(\mu_1,\cdots,\mu_i)}\xx
-\frac{\prt}{\prt
x^{(\sigma)}}\Gamma_{(\lambda,\eta)}^{(\mu_1,\cdots,\mu_i)}\xx\nn\\&+&
\delta^{(\mu_1,\cdots,\mu_i)}_{(\rho_1,\cdots,\rho_i)} \sum_{l=1}^i
\sum_{r=0}^{i-l}
\biggl(\Gamma_{(\lambda,\theta)}^{(\rho_1,\cdots,\rho_l)}\xx
\Gamma_{(\sigma,\eta)}^{(\rho_{(l+1)},\cdots,\rho_{i},\theta)}\xx-
\Gamma_{(\sigma,\theta)}^{(\rho_1,\cdots,\rho_l)}\xx
\Gamma_{(\lambda,\eta)}^{(\rho_{(l+1)},\cdots,\rho_{(i)},\theta)}\xx\biggr)\nn\\
\lbl{curvature}
\end{eqnarray}

Now the Bianchi identities (for any order $i$) follow:
\begin{eqnarray}
&&
\cD_{(\xi)}\cR_{([\lambda,\sigma],\eta)}^{(\rho_1,\cdots,\rho_i)}\xx+
\cD_{(\lambda)}\cR_{([\sigma,\xi],\eta)}^{(\rho_1,\cdots,\rho_i)}\xx+
\cD_{(\sigma)}\cR_{([\xi,\lambda],\eta)}^{(\rho_1,\cdots,\rho_i)}\xx=0\nn\\
\lbl{Bianchi}
\end{eqnarray}
and it is easy to verify
\begin{eqnarray}
&&
\cR_{([\lambda,\sigma],\eta)}^{(\rho_1,\cdots,\rho_i)}\xx+
\cR_{([\sigma,\eta],\lambda)}^{(\rho_1,\cdots,\rho_i)}\xx+\cR_{([\eta,\lambda],\sigma)}^{(\rho_1,\cdots,\rho_i)}\xx=0\nn\\
\lbl{affine}
\end{eqnarray}




In the Fang Fronsdal
 linearized  limit of Equations \rf{fangfronsdalimit},\rf{fangfronsdalimit1},\rf{fangfronsdalimit2}, the connections obey the B.R.S law
\begin{eqnarray}
\cS\Gamma_{(\nu,\sigma)}^{(\eta_1,\cdots,\eta_{j})}\xx&=&\prt_\nu\prt_\sigma\ccC^{(\eta_1,\cdots,\eta_{j})}\xx\nn\\
\end{eqnarray}

In this limit all the formalism of \cite{deWit:1979pe} can be recovered: anyway a smooth limit from our formalism to this one cannot be regained.





In the next Section we construct an higher spin Lagrangian model where the previous symmetry is realized.
\skip 1.0 cm

\section{The model construction}
\lbl{model}

This Section is devoted to the B.R.S. formulation of a Lagrangian massless Higher Spin Field Theory  model interacting with  background.
We perform here an algebraic treatment on the basis along the lines  of the Constructive Lagrangian Field Theory, in order to get a first test of this approach.
The Higher Spin Field is the main object  of our treatment, but we have to remind that the background is the special guest, which absolutely cannot be confined into a secondary role.

 We begin with a Classical Lagrangian, describing a tower of massless higher spin fields interacting with an external geometrical background, then we investigate  which conditions  allow to the highest spin component of the field to propagate and to couple with the background.

\begin{eqnarray}
&&\cL_{(Classical)}\xx= \sum_{i,j=1}^N\Biggl\{-\frac{1}{2}{\mathbb{G}}^{(\nu,\mu)}_{((\rho_1,\cdots,\rho_i),(\sigma_1,\cdots,\sigma_j))} \xx\cD_\mu\cA^{(\rho_1,\cdots,\rho_i)}\xx\cD_\nu\cA^{(\sigma_1,\cdots,\sigma_j)}\xx\nn\\%
&+& {\mathbb{F}}^{(\mu,\nu)}_{((\rho_1,\cdots,\rho_i),(\sigma_1,\cdots,\sigma_j))} \xx
\cA^{(\rho_1,\cdots,\rho_i)}\xx\sum_{k=1}^j
\cR_{([\mu,\nu],\lambda)}^{(\sigma_1,\cdots,\sigma_k)}\xx\cA^{(\lambda,\sigma_{k+1},\cdots,\sigma_j)}\xx\Biggr\}
\nn\\
\lbl{classicallagrangian}
\end{eqnarray}

In the Lagrangian in Equation \rf{classicallagrangian},the quantities ${\mathbb{G}}^{(\nu,\mu)}_{((\rho_1,\cdots,\rho_i),(\sigma_1,\cdots,\sigma_j))} \xx$ (which is symmetric under the simultaneous  exchange $\mu \longrightarrow \nu$ and $(\rho_1,\cdots,\rho_i)
 \longrightarrow (\sigma_1,\cdots,\sigma_j)$)
and ${\mathbb{F}}^{(\nu,\mu,\lambda)}_{((\rho_1,\cdots,\rho_i),(\sigma_1,\cdots,\sigma_j))} \xx$ are tensor densities and are assumed to depend on the higher order metrics.
We assume that  they obey the constraints:

 \begin{eqnarray}
&&{{\cV_{5_1}}}^{(\mu,\nu)}_{((\rho_1,\cdots,\rho_i),(\sigma_1,\cdots,\sigma_j))} \xx\equiv\cD_\mu{\mathbb{G}}^{(\nu,\mu)}_{((\rho_1,\cdots,\rho_i,\sigma_1,\cdots,\sigma_j))} \xx=0
\nn\\&&\nn\\
&&{{\cV_{5_2}}}^{(\mu,\nu)}_{((\rho_1,\cdots,\rho_i),(\sigma_1,\cdots,\sigma_j))} \xx\equiv\cD_\mu{\mathbb{F}}^{(\mu,\nu)}_{((\rho_1,\cdots,\rho_i,\sigma_1,\cdots,\sigma_j))} \xx=0
\nn\\&&\nn\\
\lbl{metricity}
\end{eqnarray}
It is easily shown that ${\mathbb{F}}^{(\mu,\nu)}_{((\rho_1,\cdots,\rho_i),(\sigma_1,\cdots,\sigma_j))} \xx$ is antisymmetric
under the  exchange $(\rho_1,\cdots,\rho_i)
 \longrightarrow (\sigma_1,\cdots,\sigma_j)$

Now; if we define the Classical Action:
\begin{eqnarray}
S_{(Classical)}=\int d^4 x \cL_{(Classical)}\xx
\end{eqnarray}

the B.R.S. invariance:
\begin{eqnarray}
\delta_{B.R.S}S_{(Classical)}=0
\end{eqnarray}

is translated, in the local form, into the Lagrangian transformation:

\begin{eqnarray}
&&\cS\cL_{(Classical)}\xx=\prt_\lambda\Biggl(\sum_{k=1}\ccC^{(\lambda)}
{\cL_{(Classical)}}\xx
\Biggr)\nn\\
&&
+\sum_{k>1}^{N-1}\prt_\lambda\Biggl(\ccC^{(\lambda,\eta_1,\cdots,\eta_k)}\xx
\Biggl(-\frac{1}{2}\biggl(\sum_{i=1}^{N-1-k}\sum_{j=1}^N
\mathbb{G}^{(\nu,\mu)}_{((\rho_1,\cdots,\rho_i;\eta_1,\cdots,\eta_k),(\sigma_1,\cdots,\sigma_j))} \xx\nn\\
&&+\sum_{i=1}^N\sum_{j=1}^{N-1-k}\mathbb{G}^{(\nu,\mu)}_{((
\rho_1,\cdots,\rho_i),(\sigma_1,\cdots,\sigma_j;\eta_1,\cdots,\eta_k))} \xx\biggr)\cD_\mu\cA^{(\rho_1,\cdots,\rho_i)}\xx\cD_\nu\cA^{(\sigma_1,\cdots,\sigma_j)}\xx
\nn\\
&+& \biggl(\sum_{i=1}^{N-1-k}\sum_{j=1}^N\mathbb{F}^{(\mu,\nu)}_{((\rho_1,\cdots,\rho_i;\eta_1,\cdots,\eta_k),(\sigma_1,\cdots,\sigma_j))} \xx+\sum_{i=1}^N\sum_{j=1}^{N-1-k}\mathbb{F}^{(\mu,\nu)}_{((\rho_1,\cdots,\rho_i
),(\sigma_1,\cdots,\sigma_j;\eta_1,\cdots,\eta_k))} \xx\biggr)\nn\\
&&
\cA^{(\rho_1,\cdots,\rho_i)}\xx\sum_{k=1}^j
\cR_{([\mu,\nu],\lambda)}^{(\sigma_1,\cdots,\sigma_k)}\xx
\cA^{(\lambda,\sigma_{k+1},\cdots,\sigma_j)}\xx\Biggr)
\nn\\
\end{eqnarray}

 which requires suitable B.R.S. transformations for the functions contained into the Lagrangian of equation \rf{classicallagrangian}. For example
\begin{eqnarray}
&&\cS{\mathbb{G}}^{(\nu,\mu)}_{((\rho_1,\cdots,\rho_i),(\sigma_1,\cdots,\sigma_j)} \xx=\prt_\lambda\Biggl(\ccC^{(\lambda)}\xx
{\mathbb{G}}^{(\nu,\mu)}_{((\rho_1,\cdots,\rho_i),(\sigma_1,\cdots,\sigma_j))} \xx\Biggr)
\nn\\&&+\sum_{m=1}^N\Biggl(\sum_{k=1}^i\prt_{\rho_k}\ccC^{(\eta_1,\cdots,\eta_m)}\xx
{\mathbb{G}}^{(\nu,\mu)}_{(\rho_1,\cdots,{}_{\widehat{\rho_k}}\cdots,\rho_i;\eta_1,\cdots,\eta_m),(\sigma_1,\cdots,\sigma_j)}\xx
\nn\\&&+\sum_{k=1}^j\prt_{\sigma_k}\ccC^{(\eta_1,\cdots,\eta_m)}\xx
{\mathbb{G}}^{(\nu,\mu)}_{((\rho_1,\cdots,\rho_i),
(\sigma_1,\cdots,{}_{\widehat{\sigma_k}}\cdots,\sigma_j;\eta_1,\cdots,\eta_m))}\xx\Biggr)
\nn\\&&-\prt_\lambda\ccC^{(\mu)}\xx
{\mathbb{G}}^{(\nu,\lambda)}_{((\rho_1,\cdots,\rho_i),(\sigma_1,\cdots,\sigma_j,
))}\xx
-\prt_\lambda\ccC^{(\nu)}\xx
{\mathbb{G}}^{(\lambda,\mu)}_{((\rho_1,\cdots,\rho_i),\eta_1(\sigma_1,\cdots,\sigma_j))}
\xx\nn\\
&&+\sum_{k>1}^{N-1}\prt_\lambda\Biggl(\ccC^{(\lambda,\eta_1,\cdots,\eta_k)}\xx
\biggl({\mathbb{G}}^{(\nu,\mu)}_{((\rho_1,\cdots,\rho_i;\eta_1,\cdots,\eta_k),(\sigma_1,\cdots,\sigma_j))} \xx+{\mathbb{G}}^{(\nu,\mu)}_{((\rho_1,\cdots,\rho_i),(\sigma_1,\cdots,\sigma_j;\eta_1,\cdots,\eta_k))} \xx\biggr)\nn\\
\lbl{sG}
\end{eqnarray}

The purpose of this section is to specify the Lagrangian, which provide  consistent equation of motion of the highest spin sector, disentangled from the  lower angular momentum components.


The general equations of motion for all tensorial fields are easily derived from the previous Lagrangian:

\begin{eqnarray}
&&\frac{\delta }{\delta \cA^{(\rho_1,\cdots,\rho_{i})}\xx}S_{(Classical)}
=\sum_{j=1}^N\Biggl[{\mathbb{G}}^{(\nu,\mu)}_{((\rho_1,\cdots,\rho_i),(\sigma_1,\cdots,\sigma_j))} \xx\Biggl(\cD_\nu\cD_\mu\Biggr) +{\mathbb{F}}^{(\nu,\mu)}_{((\rho_1,\cdots,\rho_i),
(\eta_1,\cdots,\eta_{k}))}\xx\nn\\&&
\cR^{(\eta_1,\cdots,\eta_n)}_{([\mu,\nu],\lambda)}\xx\delta^{(\lambda,\eta,_{n+1},\cdots,\eta_k)}_{(\sigma_1,\cdots,\sigma_j)}
+
{\mathbb{F}}^{(\nu,\mu)}
_{((\sigma_1,\cdots,\sigma_j),
(\eta_1,\cdots,\eta_{k}))}\xx\sum_{l=1}^k
\cR^{(\eta_1\cdots\eta_l)}_{([\mu,\nu],\lambda)}\xx\delta_{(\rho_1,\cdots,\rho_i)}^{(\lambda,\eta_{l+1},\cdots,\eta_k)}\Biggr]\cA^{(\sigma_1,\cdots,\sigma_j)}\xx\nn\\
=0\nn\\
&& i=1, \cdots N\nn\\
\lbl{equationmotion}
\end{eqnarray}

The index $i=N$ describes the evolution of the top spin component, whose
 consistency, in relation to the interaction with a background geometry, is an argument widely treated in the Literature since many years\cite{Buchdahl:1958xv}. A positive conclusion is found  in Reference \cite{Fradkin:1987ks}.

Now Equation \rf{equationmotion} has to be subjected to the constraints in Equations \rf{N-1} \rf{N-2}\rf{N-3} \rf{equationmotion}. This process presents   long calculations.

 We report it in the Appendix \rf{motionconstraints}.

 These constraints  give, a priori, links between the functions parameters $\mathbb{F}$
$\mathbb{G}$ with the curvatures and the $a\xx$ and $b\xx$ parameters.

 The results which come from these calculations allow us  to make the Statement:
\begin{Statement}
The consistency of the equation of motion of the Highest spin field is assured by precise constraints of the coupling functions which must be fixed within the Lagrangian formulation in whatever geometrical background
\lbl{Statement0}

\end{Statement}
 A more refined treatment  is reported in a sub Appendix \rf{spin2} for the spin two case,where the shorter chain constraints reveals a geometrical appeal which cannot be generalized to all the levels.


\vskip 1.0 cm




For this purpose, also on the basis of the Statement \rf{Statement0}, all these constraints, (together with condition \rf{metricity}), must be embedded within the Lagrangian formalism .

In this way we can  describe in a consistent way both the interactions of the Higher Spin  massless  particle   with the geometrical background and with other lower spin fields.

\subsection{ The  Functional method  and the Ward identities}

 First os all,due to their high tensorial content,  we cast them  in a linear, pure Slavnov variation (and zero Faddeev-Popov charged), so the invariant constraint Lagrangian term $\cL_{(Constraints)}\xx$ can be written as  :

{\footnote{A short-hand notation is extremely welcome in this occasion}}

\begin{eqnarray}
&&\cL_{(Constraints)}\xx=\cS\int d^4x\Biggl\{\sum_i\beta^{(i)}_{(\cdots)}\xx\cV_{(i)}^{(\cdots)}\xx\Biggr\}
\end{eqnarray}
(where the dots  stand for  all kind of indices for the $\cV_i$  terms)
According to a well defined technique \cite{Piguet:1984js}\cite{Piguet:1995er}, we introduce the doublet field sets $(\beta^{(i)}_{(\cdots)}\xx,\rho^{(i)}_{(\cdots)}\xx)$,  $( a_{(\cdots)}^{{i}}\xx$, $f_{(\cdots)}^{{i}}\xx)$, $( b_{(\cdots)}^{{i}}\xx$, $g_{(\cdots)}^{{i}}\xx)$ {\footnote{where $ a_{(\cdots)}^{{i}}\xx $ and $ b_{(\cdots)}^{{i}}\xx$ are the local parameters  in the $\cV_{(i)}^{(\cdots)}\xx$ constraints} }whose variations are defined as:

\begin{eqnarray}
&&\cS \beta^{(i)}_{(\cdots)}\xx=\rho^{(i)}_{(\cdots)}\xx\qquad \cS\rho^{(i)}_{(\cdots)}\xx=0\nn\\
\end{eqnarray}
\begin{eqnarray}
&&\cS a_{(\cdots)}^{{i}}\xx=f_{(\cdots)}^{{i}}\xx\qquad\cS f_{(\cdots)}^{{i}}\xx=0\nn\\
\end{eqnarray}
\begin{eqnarray}
&&\cS b_{(\cdots)}^{{i}}\xx=g_{(\cdots)}^{{i}}\xx\qquad\cS g_{(\cdots)}^{{i}}\xx=0\nn\\
\end{eqnarray}
\begin{eqnarray}
&&\cS {a^{(-1)}}^{(\cdots)}_{{i}}\xx=-\sum_{(\bullet)}{a^{(-1)}}^{(\bullet)}_{{i}}\xx f_{(\bullet)}^{{i}}\xx{a^{(-1)}}^{(\cdots)}_{{i}}\xx\nn\\
\end{eqnarray}
These variations add a new term  in the B.R.S. operator:
\begin{eqnarray}
\delta_{(Constraints)}=\int d^4 x\sum_{(\cdots)}\Biggl[ f_{(\cdots)}^{{i}}\xx\frac{\delta}{\delta a_{(\cdots)}^{{i}}\xx}+g_{(\cdots)}^{{i}}\xx\frac{\delta}{\delta b_{(\cdots)}^{{i}}\xx}+\rho^{({i})}_{(\cdots)}\xx\frac{\delta}{\delta \beta^{({i})}_{(\cdots)}\xx}\Biggr]\equiv\cQ\nn\\
\end{eqnarray}
and the total B.R.S. operator is now given:
\begin{eqnarray}
&&\delta_{(Total)}=\delta_{(B.R.S)}+\delta_{(Constraints)}\nn\\
\lbl{brstotal}
\end{eqnarray}
Consequently the total Action becomes:
\begin{eqnarray}
S_{(Total)}=S_{(Classical)}+S_{(Constraints)}
\end{eqnarray}

so that the constraint Equations \rf{metricity}, are put here as the equations of motions
\begin{eqnarray}
\frac{\delta}{\delta \rho^{(i)}_{(\cdots)}\xx}S_{(Total)}\equiv \cV_{(i)}^{(\cdots)}\xx=0\nn\\
\lbl{onshellconstraint}
\end{eqnarray}

Moreover the introduction into the total Action of the term $S_{(Constraints)}$  modifies the Equation of motion \rf{equationmotion} with terms linear in the new external fields  $\beta^{(i)}_{(\cdots)}\xx$ and $\rho^{(i)}_{(\cdots)}\xx$. This can be overcome by considering the mixed functional derivatives, for example, from Equation\rf{onshellconstraint} we can deduce:

\begin{eqnarray}
\frac{\delta^2}{\delta \rho^{(i)}_{(\cdots)}\xx \delta \cA^{(\mu_1,\cdots,\mu_N)}\yy }S_{(Total)}=0\nn\\
\lbl{onshellconstraintdouble}
\end{eqnarray}
(and similar others) which enforce the vanishing of the linear terms.

\bigskip
 Going on with our treatment, we remark that a fundamental property of $\delta_{(Total)}$ is that it  is the sum of nilpotent and anticommuting terms:

\begin{eqnarray}
&&\delta^2_{(Total)}=\delta^2_{(B.R.S)}=\delta^2_{(Constraints)}=\Biggl\{\delta_{(B.R.S)},\delta_{(Constraints)}\Biggr\}=0
\end{eqnarray}

In such a case, a famous theorem of Godement ,  can be found in Reference \cite{Godement1997bb},
states that the $\delta_{(Total)}$ cohomology  is deeply related to the one of    $\delta_{(Constraints)}$.
A field theoretic proof can be carried on using the method of Dixon \cite{Dixon:1979bs}, and the works of the Piguet group\cite{Piguet:1984js}\cite{Piguet:1995er}: if we introduce the quantities:

\begin{eqnarray}
\cQ^\dag\equiv\int d^4 x \sum_{(\cdots)}\Biggl[ a_{(\cdots)}^{{i}}\xx\frac{\delta}{\delta f_{(\cdots)}^{{i}}\xx}+b_{(\cdots)}^{{i}}\xx\frac{\delta}{\delta g_{(\cdots)}^{{i}}\xx}+\beta^{({i})}_{(\cdots)}\xx\frac{\delta}{\delta \rho^{({i})}_{(\cdots)}\xx}\Biggr]
\end{eqnarray}

and the counting operator:
\begin{eqnarray}
\cN_{Constraints}&\equiv& \int d^4 x \sum_{(\cdots)} \Biggl[a_{(\cdots)}^{{i}}\xx \frac{\delta}{\delta a_{(\cdots)}^{{i}}\xx}+
f_{(\cdots)}^{{i}}\xx \frac{\delta}{\delta f_{(\cdots)}^{{i}}\xx}+g_{(\cdots)}^{{i}}\xx\frac{\delta}{\delta g_{(\cdots)}^{{i}}\xx}\nn\\
&+&b_{(\cdots)}^{{i}}\xx\frac{\prt}{\prt b_{(\cdots)}^{{i}}\xx} +\rho^{({i})}_{(\cdots)}\xx\frac{\prt}{\prt \rho^{({i})}_{(\cdots)}\xx}
+\beta^{({i})}_{(\cdots)}\xx\frac{\delta}{\delta \beta^{({i})}_{(\cdots)}\xx}
\Biggr]\nn\\
\end{eqnarray}
we can verify that:
\begin{eqnarray}
\cN_{Constraints}=\Biggl\{\delta_{(Total)},\cQ^\dag\Biggr\}\equiv\Biggl\{\cQ,\cQ^\dag\Biggr\}
\lbl{countingauge}
\end{eqnarray}

so if $\int d^4 x\Delta^{\natural}\xx$  is an element of the cohomology of  $\delta_{(Total)}$
it is easy to derive from equation \rf{countingauge} that:
\begin{eqnarray}
\cN_{Constraints}\int d^4 x \Delta^{\natural}\xx=\delta_{(Total)}\cQ^\dag\int d^4 x \Delta^{\natural}\xx
\end{eqnarray}
which is inconsistent unless:
\begin{eqnarray}
 \cN_{Constraints}\int d^4 x \Delta^{\natural}\xx=\Biggl\{\cQ,\cQ^\dag\Biggr\}\Delta^{\natural}\xx=0\nn\\
 \lbl{parametersindepence}
 \end{eqnarray}

which says that the element of the cohomology   also belongs to the kernel of the Laplacian of $\delta_{(Constraints)}$, and so is in the $\delta_{(Constraints)}$ cohomology.

Then we derive:
\begin{eqnarray}
 \delta_{B.R.S.}\int d^4 x \Delta^{\natural}\xx=0\nn\\
 \lbl{BRSinvariance1}
 \end{eqnarray}

On this basis we enounce the Statement:
 \begin{Statement}
 If $\int d^4 x \Delta^{\natural}\xx$ in the cohomology of  $\delta_{(Total)}$, then it does not depend on the constraint parameters (Equation \rf{parametersindepence}),
 and it  is also B.R.S invariant (Equation\rf{BRSinvariance1})
 \lbl{Statement}
 \end{Statement}


This conclusion will be also fundamental for the study of our model, since states that{ \it {the physical sector is independent on the choice of the local parameters which define the highest spin extraction procedure (at least of lowest order of a perturbation expansion.
)}}.

\vskip 5mm

 Now we investigate the properties of the equations of motions. For this purpose we have to switch to the
 functional approach  starting
 with the introduction of current source terms $\cJ_{(\mu_{1}\cdots \mu_{i})}\xx$ for the quantized fields $\cA^{(\mu_{1}\cdots \mu_{i})}\xx$ ($i=1,\cdots,N$) whose B.R.S. variations can be easily inferred:
\begin{eqnarray}
&&\cS \cJ_{(\mu_{1}\cdots \mu_{i})}\xx =\sum_{l=1}^{N} \int
f^{(\rho_{1}\cdots \rho_{m})}_{[(\mu_{1}\cdots
\mu_{i}),(\nu_{1}\cdots
  \nu_{l})]}([x',x''],x) \cJ_{(\rho_{1}\cdots \rho_{m})}(x'')\ccC^{(\nu_{1}\cdots \nu_{l})}
\xp d x' d x'' \nn\\
\end{eqnarray}
more explicitly:
\begin{eqnarray}
\cS \cJ_{(\rho_1,\cdots,\rho_j)}\xx
&=& \sum_{l=1}^{N-j+1}
\Biggl[
\delta^{(\mu_1,\cdots,\mu_{l+j-1})}_{(\rho_1,\cdots,\rho_j,\sigma_2,\cdots,\sigma_l)}\prt_{\sigma_1}
\Biggl(\ccC^{(\sigma_1,\cdots,\sigma_l)}\xx\cJ_{(\mu_1,\cdots,\mu_{l+j-1})}\xx\Biggr)
\nn\\&+&\ \sum_{i=1}^{j}\prt_{(\rho_i)}\ccC^{(\sigma_1,\cdots,\sigma_l)}\xx\delta^{(\mu_1,\cdots,\mu_{l+j-1})}_{(\rho_1,\cdots
_{\widehat{(\rho_i)}},
\cdots\rho_j,\sigma_1,\cdots,\sigma_l)}\cJ_{(\mu_1,\cdots,\mu_{l+j-1})}\xx\Biggr]\nn\\
j&=&1\cdots N
\nn\\
\end{eqnarray}

in order to get:
\begin{eqnarray}
 &&
 \cS\biggl(\sum_{i=1}^{N}
\cJ_{(\mu_{1},\cdots, \mu_{i})}\xx\cA^{(\mu_{1},\cdots,
\mu_{i})}\xx\biggr) =\nn\\&&
 \sum_{j=0}^{N-1}\prt_{\lambda}\biggl(\ccC^{(\lambda,\rho_1,\cdots,\rho_j)}\xx
  \sum_{i=1}^{N-j}
   \cJ_{(\rho_1,\cdots,\rho_j,\mu_{1},\cdots ,\mu_{i})}\xx\cA^{(\mu_{1},\cdots, \mu_{i})}\xx\biggr).\nn\\
\end{eqnarray}
Notice that the top-spin current transforms only with the first order ghost, as :
\begin{eqnarray}
\cS \cJ_{(\mu_{1}\cdots \mu_{N})}\xx&=& \prt_\lambda \Biggl(\ccC^{(\lambda)}\xx \cJ_{(\mu_{1}\cdots \mu_{N})}\xx\Biggr)+\sum_{i=1}^N \prt_{\mu_i}\ccC^{(\lambda)}\xx\cJ_{(\mu_{1}\cdots,\lambda_{\widehat{\mu_i}},\cdots, \mu_{N})}\xx\nn\\
\end{eqnarray}


  We remark that the coordinate reparametrization symmetry is embodied in that part of the B.R.S. operator which is driven by the first order B.R.S. ghosts, while the part governed by the higher order ones is implied with the lowest order fields: for this reason we switch now to a Ward operator formalism.

So if we have a B.R.S. nilpotent operator $\delta_{B.R.S}$,we define the set of Ward operators as:

\begin{eqnarray}
\cW_{(\rho_1,\cdots,\rho_j)}\xx\equiv \Biggl\{\frac{\delta}{\delta \ccC^{(\rho_1,\cdots,\rho_j)}\xx},\delta_{B.R.S}\Biggr\}\quad j=1,\cdots, N-1\nn\\
\lbl{ward}
\end{eqnarray}

with local commutation relations:
\begin{eqnarray}
\Biggl[\cW_{(\nu_1,\cdots,\nu_{j})}(x'),\cW_{(\sigma_1\cdots,\sigma_{k})}(x")\Biggr]=\int d^4 x f^{(\mu_1,\cdots,\mu_{j+k-1})}_{[(\nu_1,\cdots,\nu_{{j}}),(\sigma_1\cdots,\sigma_{k})]}
(x,[x',x"])\cW_{(\mu_1,\cdots,\mu_{j+k-1})}\xx\nn\\
\end{eqnarray}
The $\cA^{(\mu_1,\cdots,\mu_j)}\xx$ fields Ward transformations are:
\begin{eqnarray}
&&\cW_{(\nu_1,\cdots,\nu_{l})}\yy\cA^{(\mu_1,\cdots,\mu_j)}\xx=\int   d^4 x" \sum_{l=1}^j f^{(\mu_1,\cdots,\mu_j)}_{[(\nu_1,\cdots,\nu_{l}),(\sigma_1\cdots,\sigma_{j-l+1})]}(x,[x',y])\cA^
{(\sigma_1,\cdots,\sigma_{j-l+1})}(x")\nn\\
&&=
\delta^{(\mu_1,\cdots,\mu_j)}_{(\rho_1,\cdots,\rho_j)}
\Biggl[\sum_{i=0}^{j-1}
\Biggl(\delta^{(\lambda,\rho_{1},\cdots,\rho_{i})}_{(\nu_1,\cdots,\nu_l)}\delta(x-y)\prt_{(\lambda)}
\cA^{(\rho_{i+1},\cdots,\rho_j)}\xx
-
\cA^{(\lambda,\rho_{1},\cdots,\rho_{i})}
\xx\prtx_{(\lambda)}\delta(x-y)
\delta^{(\rho_{i+1},\cdots,\rho_j)}_{(\nu_1,\cdots,\nu_l)}\Biggr)\Biggr]\nn\\
&&j=1,\cdots,N\quad l=1,\cdots,N-1\nn\\
\lbl{wardA}
\end{eqnarray}

 which induces an infinitesimal local transformation leading to a  rank downgrading for $l \geq 2 $. On the contrary, in the case $l=1$ we recover the usual reparametrization local Ward operator, an the tensor rank remains unchanged.

Finally it is easy to verify that for $l>j$ we have

\begin{eqnarray}
\cW_{(\nu_1,\cdots,\nu_l)}\yy\cA^{(\mu_1,\cdots,\mu_j)}\xx=0\quad l>j\nn\\
\end{eqnarray}

A fundamental remark  is that
the $\cA^{(\rho_1,\cdots,\rho_N)}\xx$ field monomial appears,
only in the Ward fields transformation of itself under the action of the reparametrization Ward operator ( that is in
 $\cW_{(\rho)}\yy\cA^{(\mu_1,\cdots,\mu_N)}\xx$)
while it is not present in any other Ward variations.

 Within the B.R.S. setting this can be seen as the consequence of the commutation relation:

\begin{eqnarray}
\Biggl[\delta_{B.R.S.},\frac{\delta}{\delta \cA^{(\rho_1,\cdots,\rho_N)}\xx}\Biggr]=\ccC^{(\lambda)}\xx\prtx_{(\lambda)} \frac{\delta}{\delta \cA^{(\rho_1,\cdots,\rho_N)}\xx}+\sum_{i=1}^n{\prtx}_{\rho_i}\ccC^{(\lambda)}\xx \frac{\delta}{\delta \cA^{(\rho_1,\cdots,\lambda_{\widehat{\rho_i}},\cdots,\rho_N)}\xx}
\nn\\
\lbl{atopinvar}
\end{eqnarray}
This equation will be relevant for our results, we introduce a hierarchy among the Ward operators  enouncing the Statement:
\begin{Statement}
Selecting the Ward operators on the basis on their action on the tensorial fields we
 call as Primary Ward operator the first order one $\cW_{(\rho)}\yy$ (which preserve the rank of the tensors), and Secondary Ward operators the others
$\cW_{(\mu_1,\cdots,\mu_l)}\yy; l=2,\cdots, N-1$ (which moves the rank from $j$ to $j-l+1$).

\end{Statement}
The Ward transformation laws of the sources terms can be easily derived:
\begin{eqnarray}
&&\cW_{(\nu_{1}\cdots \nu_{l})}\yy
 \cJ_{(\mu_{1}\cdots \mu_{i})}\xx =\sum_{l=1}^{N} \int d x''
f^{(\rho_{1}\cdots \rho_{m})}_{[(\mu_{1}\cdots
\mu_{i}),(\nu_{1}\cdots
  \nu_{l})]}([y,x''],x) \cJ_{(\rho_{1}\cdots \rho_{m})}(x'')  \nn\\
\end{eqnarray}
or, more explicitely
\begin{eqnarray}
\cW_{(\nu_1,\cdots,\nu_m)}\yy \cJ_{(\rho_1,\cdots,\rho_j)}\xx&=& \sum_{l=1}^{N-j+1}
\Biggl[
\delta^{(\mu_1,\cdots,\mu_{l+j-1})}_{(\rho_1,\cdots,\rho_j,\sigma_2,\cdots,\sigma_l)}\prt_{\sigma_1}
\Biggl(\delta_{(\nu_1,\cdots,\nu_m)}^{(\sigma_1,\cdots,\sigma_l)}\delta(x-y)
\cJ_{(\mu_1,\cdots,\mu_{l+j-1})}\xx\Biggr)
\nn\\&+&\ \sum_{i=1}^{j}\prt_{(\rho_i)}\delta(x-y)
\delta_{(\nu_1,\cdots,\nu_m)}^{(\sigma_1,\cdots,\sigma_l)}\delta^{(\mu_1,\cdots,\mu_{l+j-1})}_{(\rho_1,\cdots
_{\widehat{(\rho_i)}},
\cdots\rho_j,\sigma_1,\cdots,\sigma_l)}\cJ_{(\mu_1,\cdots,\mu_{l+j-1})}\xx\Biggr]
\nn\\
\end{eqnarray}
which induces the rank shift from $j$ to $j+m-1$, such that is immediate to derive:

\begin{eqnarray}
&&\cW_{(\nu_1,\cdots,\nu_m)}\yy \cJ_{(\rho_1,\cdots,\rho_j)}\xx=0\nn\\
&& if \quad j+m-1>N\nn\\
\end{eqnarray}

In particular the top fields sources verify:
\begin{eqnarray}
&&\cW_{(\nu_{1}\cdots \nu_{l})}\yy \cJ_{(\mu_{1}\cdots \mu_{N})}\xx=0\nn\\
&&l=2, \cdots, N-1\nn\\
\end{eqnarray}

Carrying on our model construction, and following  the standard procedure, we have to introduce the functional generator as:
\begin{eqnarray}
\cZ[\cJ]=\exp{\Biggr(-\frac{1}{\hbar} \cZ_c[{\cJ}]\Biggr)}
\end{eqnarray}
where the connected sector is given as:
\begin{eqnarray}
&&\cZ_c[{\cJ}]= \int d \cA  \Biggl[S_{Total}+\int d^4 x\biggl(\sum_{i=1}^{N}
\cJ_{(\mu_{1},\cdots, \mu_{i})}\xx\cA^{(\mu_{1},\cdots,
\mu_{i})}\xx\biggr)\nn\\
&&+\biggl(\sum_{i=1}^{N}
\gamma_{(\mu_{1},\cdots, \mu_{i})}\xx\cS\biggl(\cA^{(\mu_{1},\cdots,
\mu_{i})}\xx\biggr)\biggr)+\biggl(\sum_{i=1}^{N-1}
\zeta_{(\mu_{1},\cdots, \mu_{i})}\xx\cS\biggl(\ccC^{(\mu_{1},\cdots,
\mu_{i})}\xx\biggr)\biggr)\Biggr]\nn\\
\end{eqnarray}
and it is B.R.S. invariant:
\begin{eqnarray}
\delta \cZ_c[{\cJ}]=0 \lbl{inv0}
\end{eqnarray}

The 1-PI graphs generator is defined in a standard way as:
\begin{eqnarray}
\Gamma[\cA]&=&\Biggl[\cZ_c[{\cJ}]-\int d^4 x\biggl(\sum_{i=1}^{N}
\cJ_{(\mu_{1},\cdots, \mu_{i})}\xx\cA^{(\mu_{1},\cdots,
\mu_{i})}\xx\biggr)\Biggr]\arrowvert_{\cJ_{(\mu_{1},\cdots, \mu_{i})}=\frac{\delta \cZ_c[\cJ]}
{\delta \cA^{(\mu_{1},\cdots,
\mu_{i})}}}\nn\\
\end{eqnarray}
which, at the Classical level is B.R.S. invariant:

\begin{eqnarray}
\delta_{B.R.S.}\Gamma_{Classical}[\cA]=0
\lbl{BRSinvariance}
\end{eqnarray}

Now, the Quantum Action Principle \cite{Lam:1973qa}\cite{Gomes:1974cr}\cite{Lowenstein:1975ug}assures that:
\begin{eqnarray}
\frac{\delta}{\delta \cA^{(\rho_1,\cdots,\rho_j)}\xx}\Gamma[\cA]
=\Delta_{(\rho_1,\cdots,\rho_j)}\xx\Gamma[\cA]
\end{eqnarray}
where $\Delta_{(\rho_1,\cdots,\rho_j)}\xx$ is a local vertex insertion whose lower order expansion is given by
\begin{eqnarray}
\Delta_{(\rho_1,\cdots,\rho_j)}\xx\Gamma[\cA]=\frac{\delta}{\delta \cA^{(\rho_1,\cdots,\rho_j)}\xx}S_{Total} + O(\hbar)
\end{eqnarray}
and the first term is a  fields polynomial given by the classical equation of motions.

Now from Equations \rf{wardA} we can (again) verify that we have a hierarchy of commutation properties between Ward operators and fields function derivatives:

\begin{eqnarray}
&&\Biggl[\frac{\delta}{\delta \cA^{(\rho_1,\cdots,\rho_{N-k})}\xx},\cW_{(\sigma_1,\cdots,\sigma_j)}\yy\Biggr]=0
\quad j-k-1>0 \nn\\
&&k=0,\cdots, N-1\nn\\
&&j=1,\cdots,N-1\nn\\
\lbl{commzero}
\end{eqnarray}

 from which we can derive that all the Secondary Ward operators commute with the functional derivative with respect the highest tensor spin field.
With the increasing of $k$, the number  the number of commuting Ward operators decrease.


 Notice that we must underline that highest order ${\mathbb{G}}^{(\nu,\mu)}_{((\rho_1,\cdots,\rho_N),(\sigma_1,\cdots,\sigma_N)} \xx,$ ${\mathbb{F}}^{[\nu,\mu]}_{((\rho_1,\cdots,\rho_N),(\sigma_1,\cdots,\sigma_N)} \xx,$fields functions B.R.S. variations depend only on the order one ghost $\ccC^{(\mu)}\xx$, so that in terms of the Secondary Ward operators we also derive:

\begin{eqnarray}
&&\cW_{(\theta_1,\cdots,\theta_k)} \yy{\mathbb{G}}^{(\nu,\mu)}_{((\rho_1,\cdots,\rho_N),(\sigma_1,\cdots,\sigma_N)} \xx=0;\nn\\
&&\cW_{(\theta_1,\cdots,\theta_k)}\yy{\mathbb{F}}^{(\nu,\mu)}_{((\rho_1,\cdots,\rho_N),
(\sigma_1,\cdots,\sigma_N))} \xx=0\nn\\
&&k=2,\cdots N-1\nn\\
\end{eqnarray}

and similarly for the others external fields in the Lagrangian \rf{classicallagrangian}.

Now, equation \rf{BRSinvariance} gives, in a Ward operator language the identities:
\begin{eqnarray}
&&\cW_{(\sigma_1,\cdots,\sigma_k)}\yy S_{(Classical)}=0\nn\\
\lbl{wardactioninvariance1}
\end{eqnarray}
and:
\begin{eqnarray}
&&\cW_{(\sigma_1,\cdots,\sigma_k)}\yy S_{(Constraints)}=Order\Biggl\{\cA^{(\mu_1,\cdots,\mu_j)}\yy\Biggr\}
k=2,\cdots N-1\qquad j<N\nn\\
\lbl{wardactioninvariance2}
\end{eqnarray}
(which is due to the fact that: $\cW_{(\sigma_1,\cdots,\sigma_k)}\yy\ccC^{(\mu)}\xx=0,k=2,\cdots,N-1$)

and the previous Equations\rf{wardactioninvariance1} \rf{wardactioninvariance2} imply:
\begin{eqnarray}
&&\frac{\delta}{\delta \cA^{(\rho_1,\cdots,\rho_N)}\xx}\cW_{(\sigma_1,\cdots,\sigma_k)}\yy S_{(Total)}=0\nn\\
&&k=2,\cdots,N-1\nn\\
\lbl{totalcommutator}
\end{eqnarray}
 so we can state:

 \begin{Result}
From Equations \rf{commzero} and \rf{totalcommutator} we derive:

\begin{eqnarray}
\cW_{(\sigma_1,\cdots,\sigma_k)}\yy\Biggl[\frac{\delta }{\delta \cA^{(\rho_1,\cdots,\rho_N)}\xx} S_{(Total)}
\Biggr]=0\quad k\geq 2\nn\\
\lbl{motioninvariance}
\end{eqnarray}
which means that
 the $\cA^{(\rho_1,\cdots,\rho_N)}\xx$ Classical equations of motions are invariant under the infinitesimal transformations  induced by the Secondary Ward operators $\cW_{(\sigma_1,\cdots,\sigma_k)}\yy,k=2,\cdots,N-1$

This conclusion is valid for the evolution of all spin components of the tensor field.
\lbl{Result}
\end{Result}

\vskip 0.5cm
Our previous result  generalizes   to all  n - points 1-PI irreducible Green functions
\begin{eqnarray}
\Biggl[\prod_{i=1}^n\frac{\delta }{\delta \cA^{(\rho^{(i)}_1,\cdots,\rho^{(i)}_N)}(x_i)}\Biggr]\cS_{(Total)
}
\end{eqnarray}
 we can improve it (at the Classical level) from the Equations \rf{wardactioninvariance1} and \rf{commzero}  as:
\begin{eqnarray}
&&0=\Biggl[\cW_{(\sigma_1,\cdots,\sigma_k)}\yy,
\prod_{i=1}^n\frac{\delta}{\delta \cA^{(\rho^{(i)}_1,\cdots,\rho^{(i)}_N)}(x_i)}\Biggr]
\cS_{(Total)}\nn\\
&&\equiv \cW_{(\sigma_1,\cdots,\sigma_k)}\yy
\Biggl[\prod_{i=1}^n\frac{\delta}{\delta \cA^{(\rho^{(i)}_1,\cdots,\rho^{(i)}_N)}(x_i)}\Biggr]\cS_{(Total)}=
0\quad \forall n=1,2,\cdots\quad k= 2,\cdots,N-1\nn\\
\lbl{gaugeinvariancenpoint}
\end{eqnarray}

and the further inclusion of the restriction due to the Equations \rf{N-1},\rf{N-2},\rf{N-3}\rf{Lorentzconditionmotion} lower the
number of commuting Ward operators as before.

So we are ready to sum up our result which selects the symmetry which rules both the highest spin equation of motion and the ones of the constraints which pick the top spin component. So we collect our results into:

\vskip 1.5 cm





The previous results are obtained at the Classical level.However, if we want to explore their   Quantum Extension we have to refer the reader to Reference \cite{Bandelloni:2011zz} where
 cohomology of our algebra is studied with the aid of the spectral sequences technique.

The most important conclusion of this study (for all the details see the Appendix of the previous reference)  is that the anomalies  are confined within the algebra carried by the Primary Ward operator (diffeomorphism sector).
So, besides the usual Gravitational Anomaly \cite{Alvarez-Gaume:1983ig}, no other anomaly  appears  at the first Quantum meaningful  perturbative  order. So at this level the results given in \rf{Result} hold their validity.

\vskip 1.0 cm

\section{Conclusions}
\lbl{conclusions}

This paper can be considered as the second chapter of the Higher-Spin approach begun in reference \cite{Bandelloni:2011zz}.
We have studied here the structure, symmetries and consequences of the equations of motion of the highest spin field: up to this point, we have performed an operator  and algebraic approach supported by the basilar Theorems of Relativistic Quantum Field Theory.

The next step to  could be now to prepare a method which admits an affordable calculations program to get  touchable and reliable physical results.

The most promising way  is, in our opinion, to use extensively the Ward identities and to concentrate on the background Lagrangian which we have not analyzed here.

\vskip  3.0cm
\centerline{\Large{\bf{Acknowledgement}}}

Clarifying discussions with Alberto Blasi are gratefully acknowledged.
\vskip 1.0 cm
\appendix
\centerline{\Large{\bf{Appendices}}}
\vskip 1.0 cm
\section{The extended derivative formalism}
\lbl{extendederivative}

In this technical appendix we want to refine aspects due to the introduction of  the extended derivative notation defined in Equation\rf{derextended}.

 We want first to point out    an unexpected and
quite useful shortcut is taken by writing the B.R.S. operator
$\delta$ in the Fock space\cite{Dixon:1979bs}(where the fields and
their derivatives are considered as independent objects) and
introducing the extended derivative  operators (we shall call them
$\prt$-operators):
This  leads to a great reduction of tedious calculations, and we
can take advantage of a solution found already in the
literature\cite{Bandelloni:1988ws}.

It is straightforward to derive from the very definition the
general expressions:
\begin{eqnarray}
\prt_{(\nu_1,\cdots,\nu_m)}
\left\{\begin{array}{c}\ccC^{(\mu_1\cdots,\mu_j)}\xx\\
\ccA^{(\mu_1\cdots,\mu_j)}\xx\\
\end{array}\right.&=&
\delta^{(\mu_1,\cdots,\mu_j)}_{(\rho_1,\cdots,\rho_j)}
\delta^{(\lambda,\rho_{j-m+2},\cdots,\rho_j)}_{(\nu_1,\cdots,\nu_m)}
\prt_{(\lambda)}\left\{\begin{array}{c}
\ccC^{(\rho_1,\cdots,\rho_{j-m+1})}\xx\nn\\
\ccA^{(\rho_1,\cdots,\rho_{j-m+1})}\xx\\
\end{array}\right.\\
{\mbox{for }}&& m\leq j,\\
\prt_{(\nu_1,\cdots,\nu_l)}\ccA_{(\rho_1\cdots,\rho_j)}\xx
&=&\delta_{(\nu_1,\cdots,\nu_l)} ^{(\sigma_1,\cdots,\sigma_l)}
\delta^{(\mu_1,\cdots,\mu_{l+j-1})}_{(\rho_1,\cdots,\rho_j,\sigma_2,\cdots,\sigma_l)}
\prt_{\sigma_1}\ccA_{(\mu_1,\cdots,\mu_{l+j-1})}\xx\nn\\
\lbl{derhig}
\end{eqnarray}
\begin{eqnarray}
\prt_{(\nu_1,\cdots,\nu_m)}
\left\{\begin{array}{c}\ccC^{(\mu_1\cdots,\mu_j)}\xx\nn\\
\ccA^{(\mu_1\cdots,\mu_j)}\xx\\
\end{array}\right.&=&0, \qquad \mbox{for} \quad m> j .\nn\\
\prt_{(\nu_1,\cdots,\nu_l)}\ccA_{(\rho_1\cdots,\rho_j)}\xx&=&0,  \qquad \mbox{for} \quad l>N-j+1.\nn\\
\lbl{derhig0}
\end{eqnarray}
If we define the grading of these fields as the order of their
symmetric indices, then the operators
$\prt_{(\nu_1,\cdots,\nu_i)}$ map a monomial
$\ccA^{(\mu_1\cdots,\mu_j)}\xx$ field of grading $-j$ to the
derivative of this field with (in general) a different grading
$m-j-1$. Note that for $m=1$ we recover the familiar derivative
operator \cite{Bandelloni:1985ue}.

An essential remark is that, for the upper indices fields, for
$j=m$, we get a symmetrized combination of derivatives of the
first level fields; a fact  that will be relevant for the sequel.

On the other hand, the index contraction procedure
 cannot be performed in a naive way:

 for example for $m<j$:
\begin{eqnarray}
\prt_{(\mu_1,\cdots,\mu_m)}
\left\{\begin{array}{c}\ccC^{(\mu_{1}\cdots,\mu_j)}\xx\\
\ccA^{(\mu_1\cdots,\mu_j)}\xx\\
\end{array}\right.=
\prt_{(\lambda)}\left\{\begin{array}{c}\ccC^{(\lambda,\mu_{m+1}\cdots,\mu_j)}\xx\\
\ccA^{(\lambda,\mu_{m+1}\cdots,\mu_j)}\xx\\
\end{array}\right.
\nn\\
\lbl{dertrace}
\end{eqnarray}

so, in particular, the total contraction:
\begin{eqnarray}
\prt_{(\mu_1,\cdots,\mu_j)}
\left\{\begin{array}{c}\ccC^{(\mu_1\cdots,\mu_j)}\xx\\
\ccA^{(\mu_1\cdots,\mu_j)}\xx\\
\end{array}\right.&=&\delta_{(\mu_1,\cdots,\mu_j)}^{(\mu_1,\cdots,\mu_{j-1}, \lambda)}
 \prt_{(\lambda)}\left\{\begin{array}{c}
\ccC^{(\mu_j)}\xx\\
\ccA^{(\mu_j)}\xx\\
\end{array}\right.\nn\\
&=&
\prt_{(\lambda)}
\left\{\begin{array}{c}
\ccC^{(\lambda)}\xx\\
\ccA^{(\lambda)}\xx\\
\end{array}\right.\nn\\
\lbl{dertracehig}
\end{eqnarray}

The properties of the operators in Eq. \rf{derextended} can be
summarized as follows:

{ {
  \begin{itemize}
  \item The $\prt$-operators are invariant under $\delta$:
    \begin{eqnarray}
      \Biggl[\prt_{(\mu_1,\cdots,\mu_j)},\delta
      \Biggr]=\Biggl[\Biggl\{\frac{\prt}{\prt \ccC^{{(\mu_1\cdots,
            \mu_j)}}\xx},\delta\Biggr\},\delta \Biggr]=0
            \lbl{commd}
    \end{eqnarray}
  \item The $\prt$-operators commute with the underived ghosts derivatives
    in the Fock space:
    \begin{eqnarray}
      \Biggl[\prt_{(\nu_1,\cdots,\nu_i)},
      \frac{\prt}{\prt \ccC^{{(\mu_1\cdots,\mu_j)}}\xx}\Biggr]=0\nn\\
      \lbl{commC}
    \end{eqnarray}
  \item The $\prt$-operators commute among themselves:
    \begin{eqnarray}
      \Biggl[\prt_{(\nu_1,\cdots,\nu_k)},\prt_{(\mu_1,
        \cdots,\mu_j)} \Biggr]=\Biggl[\prt_{(\nu_1,\cdots,\nu_k)},
      \Biggl\{\frac{\prt}{\prt \ccC^{{(\mu_1\cdots,\mu_j)}}\xx},\delta
      \Biggr\} \Biggr]=0\nn\\
      \lbl{p1}
    \end{eqnarray}
  \item The $\prt$-operators of a product of fields obeys the usual rule:
\begin{eqnarray}
&&\prt_{(\nu_1,\cdots,\nu_k)}\Biggl[\cO^{(\mu_1,\cdots,\mu_j)}_{(\rho_1,\cdots,\rho_i)}\xx
\ccM^{(\sigma_1,\cdots,\sigma_l)}_{(\eta_1,\cdots,\eta_m)}\xx\Biggr]=\Biggl\{\frac{\prt}{\prt
\ccC^{{(\nu_1\cdots,
\nu_k)}}\xx},\delta\Biggr\}\Biggl[\cO^{(\mu_1,\cdots,\mu_j)}_{(\rho_1,\cdots,\rho_i)}\xx
\ccM^{(\sigma_1,\cdots,\sigma_l)}_{(\eta_1,\cdots,\eta_m)}\xx\Biggr]\nn\\
&&=\Biggl[\prt_{(\nu_1,\cdots,\nu_k)}
\cO^{(\mu_1,\cdots,\mu_j)}_{(\rho_1,\cdots,\rho_i)}\xx
\Biggr]\ccM^{(\sigma_1,\cdots,\sigma_l)}_{(\eta_1,\cdots,\eta_m)}\xx
+\cO^{(\mu_1,\cdots,\mu_j)}_{(\rho_1,\cdots,\rho_i)}\xx\Biggl[\prt_{(\nu_1,\cdots,\nu_k)}
\ccM^{(\sigma_1,\cdots,\sigma_l)}_{(\eta_1,\cdots,\eta_m)}\xx\Biggr]\nn\\
\lbl{product}
\end{eqnarray}
for whatever Faddeev-Popov uncharghed monomials
$\cO^{(\mu_1,\cdots,\mu_j)}_{(\rho_1,\cdots,\rho_i)}\xx$,
$\ccM^{(\sigma_1,\cdots,\sigma_l)}_{(\eta_1,\cdots,\eta_m)}\xx$.
 \end{itemize}
}}

We  remark that the result of operating  with
$\prt_{(\mu_1,\cdots,\mu_j)}$ depends on the tensor order of the
object on which it acts: its definition \rf{derextended} is {\it
intrinsically} defined in the Fock space, where the derivatives
are always ``pasted'' to the fields.

{\it So we must must be cautious  with the
$\prt_{(\mu_1,\cdots,\mu_k)}$-operators and to use them (for
$k\geq 2$) as common derivative operators, in the sense that they
{\it do not} transform a field into a derivative of the {\it same}
field. A direct check can be done from Eq \rf{product}: a
$\prt$-derivative ( with order greater than two) of a fields
product
 is not an ordinary total  derivative.
 }

\vspace{0.25 cm}
\section{The Equation of motion constraints}
\lbl{motionconstraints}

We now rephrase it within our framework,  embedding it within   the conditions of Equations \rf{N-1},\rf{N-2},\rf{N-3}\rf{Lorentzconditionmotion}. More specifically we repeat  (for the massless situation) the treatment just seen in the Introduction. So, in order to isolate
the highest helicity state,we fix the constraints:

\begin{eqnarray}
&&\cD^{(\rho_{i}))}\frac{\delta }{\delta \cA^{(\rho_1,\cdots,\rho_{i})}\xx}S_{(Classical)}
=\cD^{(\rho_{i}))}\sum_{j=1}^N\Biggl[{\mathbb{G}}^{(\nu,\mu)}_{((\rho_1,\cdots,\rho_i),(\sigma_1,\cdots,\sigma_j))} \xx\Biggl(\cD_\nu\cD_\mu\Biggr) +\sum_{k=1}^i{\mathbb{F}}^{(\nu,\mu)}_{((\rho_1,\cdots,\rho_i),
(\eta_1,\cdots,\eta_{k}))}\xx\nn\\&&
\cR^{(\eta_1,\cdots,\eta_n)}_{([\mu,\nu],\lambda)}\xx\delta^{(\lambda,\eta,_{n+1},\cdots,\eta_k)}_{(\sigma_1,\cdots,\sigma_j)}
+
\sum_{k=1}^i{\mathbb{F}}^{(\nu,\mu)}
_{((\sigma_1,\cdots,\sigma_j),
(\eta_1,\cdots,\eta_{k}))}\xx\sum_{l=1}^k
\cR^{(\eta_1\cdots\eta_l)}_{([\mu,\nu],\lambda)}\xx\delta_{(\rho_1,\cdots,\rho_i)}
^{(\lambda,\eta_{l+1},\cdots,\eta_k)}\Biggr]\cA^{(\sigma_1,\cdots,\sigma_j)}\xx
=0\nn\\
&& i=1, \cdots N\nn\\
\nn\\
\lbl{Lorentzequation}
\end{eqnarray}
First, to get the transversality conditions we fix:
\begin{eqnarray}
&&{{\cV_6}_m}^{(\{\nu,\mu\})}_{((\rho_1,\cdots,\rho_{i-1}),(\sigma_1,\cdots,\sigma_j))} \xx\equiv \cD^{(\rho_i)}{\mathbb{G}_j}^{(\{\nu,\mu\})}_{((\rho_1,\cdots,\rho_i),(\sigma_1,\cdots,\sigma_j))} \xx=0 \quad m=1,2\nn\\
&&{\cV_7}^{(\nu,\mu)}_{((\rho_1,\cdots,\rho_{i-1}),
(\eta_1,\cdots,\eta_{k}))}\xx\equiv\cD^{(\rho_i)}{\mathbb{F}}^{(\nu,\mu)}_{((\rho_1,\cdots,\rho_i),
(\eta_1,\cdots,\eta_{k}))}\xx=0\nn\\
&&{\cV_8}_{((\sigma_1,\cdots,\sigma_j),
(\rho_1,\cdots,\rho_{i-1}))}\xx\equiv\cD^{(\rho_i)}\Biggl[{\mathbb{F}}^{(\nu,\mu)}
_{((\sigma_1,\cdots,\sigma_j),
(\eta_1,\cdots,\eta_{k}))}\xx
\delta_{(\rho_1,\cdots,\rho_i)}^{(\lambda,\eta_{l+1},\cdots,\eta_k)}\Biggr]=0\nn\\
&&i,j,l,k=1,\cdots,N;\nn\\
\lbl{transversality}
\end{eqnarray}
So we get:
\begin{eqnarray}
&&g^{(\rho_i,\lambda)}\xx\sum_{j=1}^N\Biggl[{\mathbb{G}}^{(\nu,\mu)}_{((\rho_1,\cdots,\rho_i),(\sigma_1,\cdots,\sigma_j))} \xx\biggl(\cR_{([\lambda,\nu]\mu)}^\eta\cD_\eta\cA^{(\sigma_1,\cdots,\sigma_j)}\xx+
\nn\\&&+\cD_{\nu}\biggl(
\sum_{k=1}^j\cR_{([\lambda,\mu],\eta)}^{(\sigma_1,\cdots,\sigma_k)}\cA^{(\eta,\sigma_{k+1},\cdots,\sigma_j)}\xx\biggr)
+\Biggl(\cD_\nu\cD_\mu\Biggr) \cD_\lambda \cA^{(\sigma_1,\cdots,\sigma_j)}\xx\biggr)\nn\\
&&+{\mathbb{F}}^{(\nu,\mu)}_{((\rho_1,\cdots,\rho_i),
(\eta_1,\cdots,\eta_{k}))}\xx
\cD_\lambda\biggl(\cR^{(\eta_1,\cdots,\eta_n)}_{([\mu,\nu],\xi)}\xx
\delta^{(\xi,\eta,_{n+1},\cdots,\eta_k)}_{(\sigma_1,\cdots,\sigma_j)}
\cA^{(\sigma_1,\cdots,\sigma_j)}\xx\biggr)\nn\\&&
+
{\mathbb{F}}^{(\nu,\mu)}
_{((\sigma_1,\cdots,\sigma_j),
(\eta_1,\cdots,\eta_{k}))}\xx\sum_{l=1}^k
\delta_{(\rho_1,\cdots,\rho_i)}^{(\xi,\eta_{l+1},\cdots,\eta_k)}\cD_\lambda\biggl(\cR^{(\eta_1\cdots\eta_l)}_{([\mu,\nu],\xi)}\xx\cA^{(\sigma_1,\cdots,\sigma_j)}\xx\biggr)\Biggr]
\nn\\
\lbl{Lorentzz}
&&=0\nn\\
\end{eqnarray}
The triple derivative term in the fields is zero if we use the Lorentz condition and   we fix:

\begin{eqnarray}
g^{(\rho_i,\lambda)}\xx
{\mathbb{G}}^{(\nu,\mu)}_{((\rho_1,\cdots,\rho_i),(\sigma_1,\cdots,\sigma_j))}\xx
=\delta^\lambda_{\sigma_j}{\mathbb{H}}^{(\nu,\mu)}_{((\rho_1,\cdots,\rho_{i-1})
,(\sigma_1,\cdots,\sigma_{j-1}))}\xx\nn\\
\lbl{triple}
\end{eqnarray}
which implies:
\begin{eqnarray}
{\mathbb{G}}^{(\nu,\mu)}_{((\rho_1,\cdots,\rho_i),(\sigma_1,\cdots,\sigma_j))}\xx=
{g^{(-1)}}_{(\rho_i,\sigma_j)}\xx
{\mathbb{H}}^{(\nu,\mu)}_{((\rho_1,\cdots,\rho_{i-1})
,(\sigma_1,\cdots,\sigma_{j-1}))}\xx
\end{eqnarray}

so we get:

\begin{eqnarray}
&&\Biggl[g^{(\rho_i,\lambda)}\xx
{\mathbb{G}}^{(\nu,\mu)}_{((\rho_1,\cdots,\rho_i),(\sigma_1,\cdots,\sigma_j))}\xx
\biggl(\cR_{([\lambda,\nu]\mu)}^\eta\cD_\eta\cA^{(\sigma_1,\cdots,\sigma_j)}\xx
+\cD_{\nu}\biggl(
\sum_{k=1}^j\cR_{([\lambda,\mu],\eta)}^{(\sigma_1,\cdots,\sigma_k)}\cA^{(\eta,\sigma_{k+1},\cdots,\sigma_j)}\xx\biggr)
\biggr)\nn\\
&&+g^{(\rho_i,\lambda)}\xx{\mathbb{F}}^{(\nu,\mu)}_{((\rho_1,\cdots,\rho_i),
(\eta_1,\cdots,\eta_{k}))}\xx
\cD_\lambda\biggl(\cR^{(\eta_1,\cdots,\eta_n)}_{([\mu,\nu],\xi)}\xx
\delta^{(\xi,\eta,_{n+1},\cdots,\eta_k)}_{(\sigma_1,\cdots,\sigma_j)}
\cA^{(\sigma_1,\cdots,\sigma_j)}\xx\biggr)\nn\\&&
+
g^{(\rho_i,\lambda)}\xx{\mathbb{F}}^{(\nu,\mu)}
_{((\sigma_1,\cdots,\sigma_j),
(\eta_1,\cdots,\eta_{k}))}\xx\sum_{l=1}^k
\delta_{(\rho_1,\cdots,\rho_i)}^{(\xi,\eta_{l+1},\cdots,\eta_k)}\cD_\lambda\biggl(\cR^{(\eta_1\cdots\eta_l)}_{([\mu,\nu],\xi)}\xx\cA^{(\sigma_1,\cdots,\sigma_j)}\xx\biggr)\Biggr]
\nn\\
\lbl{Lorentzz1}
&&=0\nn\\
\end{eqnarray}

So we are left with two kind of terms: one which contains the underived tensor fields, and the second with first order derivative terms $\cD_\xi\cA^{(\kappa_1,\cdots,\kappa_r)}\xx$.

Anyway writing the first type of contributions:
\begin{eqnarray}
&&\sum_{j=1}^N\Biggl[
g^{(\rho_i,\lambda)}\xx
{\mathbb{G}}^{(\nu,\mu)}_{((\rho_1,\cdots,\rho_i),(\sigma_1,\cdots,\sigma_j))}\xx
\biggl(
\sum_{k=1}^j\cD_{\nu}\cR_{([\lambda,\mu],\eta)}^{(\sigma_1,\cdots,\sigma_k)}
\delta^{(\eta,\sigma_{k+1},\cdots,\sigma_j)}
_{(\kappa_1,\cdots,\kappa_r)}
\biggr)\nn\\
&&+g^{(\rho_i,\lambda)}\xx{\mathbb{F}}^{(\nu,\mu)}_{((\rho_1,\cdots,\rho_i),
(\eta_1,\cdots,\eta_{k}))}\xx
\biggl(\sum_{n=1}^k\cD_\lambda\cR^{(\eta_1,\cdots,\eta_n)}_{([\mu,\nu],\xi)}\xx
\delta^{(\xi,\eta,_{n+1},\cdots,\eta_k)}_{(\kappa_1,\cdots,\kappa_r)}\biggr)
\nn\\
&&+g^{(\rho_i,\lambda)}\xx
{\mathbb{F}}^{(\nu,\mu)}
_{((\sigma_1,\cdots,\sigma_j),
(\eta_1,\cdots,\eta_{k}))}\xx\sum_{l=1}^k
\delta_{(\rho_1,\cdots,\rho_i)}^{(\xi,\eta_{l+1},\cdots,\eta_k)}
\biggl(\cD_\lambda\cR^{(\eta_1\cdots\eta_l)}_{([\mu,\nu],\xi)}\xx
\Biggr)
\Biggr]\cA^{(\kappa_1,\cdots,\kappa_r)}\xx
\nn\\
&&=0\nn\\
\lbl{A}
\end{eqnarray}

we discover that all the terms in the square brackets in Equation \rf{A} are zero,
due to Bianchi Equations \rf{Bianchi}imposing  for  ${\mathbb{G}}$ and ${\mathbb{F}}$ the links:

\begin{eqnarray}
g^{(\rho_i,\lambda)}\xx\left(\begin{array}{c}{\mathbb{F}}^{(\nu,\mu)}_{((\rho_1,\cdots,\rho_i),
(\eta_1,\cdots,\eta_l))}\xx\\
    \\
{\mathbb{G}}^{(\nu,\mu)}_{((\rho_1,\cdots,\rho_i),
(\eta_1,\cdots,\eta_l))}\xx\end{array}\right)=g^{(\rho_i,\nu)}\xx
\left(\begin{array}{c}{\mathbb{F}}^{(\mu,\lambda)}_{((\rho_1,\cdots,\rho_i),
(\eta_1,\cdots,\eta_l))}\xx\\
    \\
{\mathbb{G}}^{(\mu,\lambda)}_{((\rho_1,\cdots,\rho_i),
(\eta_1,\cdots,\eta_l))}\xx\end{array}\right)
=g^{(\rho_i,\mu)}\xx
\left(\begin{array}{c}{\mathbb{F}}^{(\lambda,\nu)}_{((\rho_1,\cdots,\rho_i),
(\eta_1,\cdots,\eta_l))}\xx\\
    \\
{\mathbb{G}}^{(\lambda,\nu)}_{((\rho_1,\cdots,\rho_i),
(\eta_1,\cdots,\eta_l))}\xx\end{array}\right)
\nn\\
\lbl{cyclic}
\end{eqnarray}

so for ${\mathbb{F}}$ we choose:
\begin{eqnarray}
{\cV_{9a}}^{(\nu,\mu)}_{((\rho_1,\cdots,\rho_i),
(\sigma_1,\cdots,\sigma_j))}\xx\equiv{\mathbb{F}}^{(\nu,\mu)}_{((\rho_1,\cdots,\rho_i),
(\sigma_1,\cdots,\sigma_j))}\xx=g^{(\eta,\mu)}\xx g^{(\tau,\nu)}\xx{\mathfrak{f}}_{((\rho_1,\cdots,\rho_i,[\eta,\tau]),
(\sigma_1,\cdots,\sigma_j))}\xx\nn\\
\lbl{ourF}
\end{eqnarray}
and the special symbol ${\mathfrak{f}}$ specifies that is a density (with weight one) instead of a tensor!

and:
\begin{eqnarray}
{\cV_{9b}}^{(\nu,\mu)}_{((\rho_1,\cdots,\rho_{i})
,(\sigma_1,\cdots,\sigma_{j}))}\xx\equiv{\mathbb{G}}^{(\nu,\mu)}_{((\rho_1,\cdots,\rho_{i})
,(\sigma_1,\cdots,\sigma_{j}))}\xx
&=&{g^{(-1)}}_{(\rho_{i},\sigma_{j})}\xx g^{(\mu,\eta)}\xx g^{(\nu,\kappa)}\xx\Biggl[{\mathfrak{g}}_{((\rho_1,\cdots,\rho_{i-1}),(\sigma_1,\cdots,\sigma_{j-1}),(\eta,\kappa))}\Biggr]
\nn\\
\lbl{ourGn}
\end{eqnarray}

and we are left with the term:
\begin{eqnarray}
&&\sum_{j=1}^N\Biggl[
g^{(\rho_i,\lambda)}\xx
{\mathbb{G}}^{(\nu,\mu)}_{((\rho_1,\cdots,\rho_i),(\sigma_1,\cdots,\sigma_j))}\xx
\biggl(\cR_{([\sigma_j,\nu]\mu)}^\eta\delta^{(\sigma_1,\cdots,\sigma_j)}_{(\kappa_1,\cdots,\kappa_r)}
\delta_\eta^\xi+\biggl(
\sum_{k=1}^j
\cR_{([\sigma_j,\mu],\eta)}^{(\sigma_1,\cdots,\sigma_k)}\delta^{(\eta,\sigma_{k+1},\cdots,\sigma_j)}
_{(\kappa_1,\cdots,\kappa_r)}\delta_\nu^\xi
\biggr)
\biggr)\nn\\
&&+g^{(\rho_i,\lambda)}\xx{\mathbb{F}}^{(\nu,\mu)}_{((\rho_1,\cdots,\rho_i),
(\eta_1,\cdots,\eta_{k}))}\xx
\biggl(
\sum_{n=1}^k\cR^{(\eta_1,\cdots,\eta_n)}_{([\mu,\nu],\upsilon)}\xx
\delta^{(\upsilon,\eta,_{n+1},\cdots,\eta_k)}_{(\kappa_1,\cdots,\kappa_r)}
\delta_\lambda^\xi\biggr)\nn\\&&
+g^{(\rho_i,\lambda)}\xx
{\mathbb{F}}^{(\nu,\mu)}
_{((\sigma_1,\cdots,\sigma_j),
(\eta_1,\cdots,\eta_{k}))}\xx\sum_{l=1}^k
\biggl(
\cR^{(\eta_1\cdots\eta_l)}_{([\mu,\nu],\upsilon)}\xx
\delta_{(\rho_1,\cdots,\rho_i)}^{(\upsilon,\eta_{l+1},\cdots,\eta_k)}\delta^{(\sigma_1,\cdots,\sigma_j)}_{(\kappa_1,\cdots,\kappa_r)}
\delta_\lambda^\xi\biggr)\Biggr]\cD_\xi\cA^{(\kappa_1,\cdots,\kappa_r)}\xx
\nn\\
&&=0\nn\\
\lbl{DA}
\end{eqnarray}

By the way, with the previous assumptions the term $g^{(\rho_i,\lambda)}\xx
{\mathbb{G}}^{(\nu,\mu)}_{((\rho_1,\cdots,\rho_i),(\sigma_1,\cdots,\sigma_j))}\xx\cR_{([\sigma_j,\nu]\mu)}^\eta\delta^{(\sigma_1,\cdots,\sigma_j)}_{(\kappa_1,\cdots,\kappa_r)}
\delta_\eta^\xi $ is zero  using the condition \rf{affine}.
So, for whatever fields value we must rewrite, using Equation \rf{cyclic}, the  Equation\rf{DA} into:

\begin{eqnarray}
&&
{\cV_{9c}}^{((\lambda,\xi),(\theta,\tau,\chi))}_{((\rho_1,\cdots,\rho_{i}),(\eta_{1},\cdots,\eta_l)
,(\kappa_1,\cdots,\kappa_r))}\xx\equiv
\Biggl[
{\mathbb{G}}^{(\{\xi,\tau\})}_{((\rho_1,\cdots,\rho_i),(\eta_1,\cdots,\eta_k))}\xx
\delta^{(\chi,\eta_{l+1},\cdots,
\eta_k)}
_{(\kappa_1,\cdots,\kappa_r)}
+
\biggl(
{\mathbb{F}}^{([\xi,\tau])}_{((\rho_1,\cdots,\rho_i),
(\eta_1,\cdots,\eta_{k}))}\xx
\delta^{(\chi,\eta,_{l+1},\cdots,\eta_k)}_{(\kappa_1,\cdots,\kappa_r)}
\nn\\&&
+
{\mathbb{F}}^{([\xi,\tau])}
_{((\kappa_1,\cdots,\kappa_r),
(\eta_1,\cdots,\eta_{k}))}\xx
\delta_{(\rho_1,\cdots,\rho_i)}^{(\chi,\eta_{l+1},\cdots,\eta_k)}
\biggr)
\Biggr]\cR_{([\theta,\tau],\chi)}^{(\eta_{1},\cdots,\eta_l)}\xx
=0\nn\\
\lbl{condtransv1}
\end{eqnarray}

 Conditions  \rf{condtransv1} imply (a priori) the interplay between  ${\mathbb{F}}$ ${\mathbb{G}}$  and the curvatures.If we require the independence of our analysis from the geometry, then we have to require that each square bracket content of Equations \rf{condtransv1} will be identically zero. This links the ${\mathbb{G}}$ functions to a symmetric combination of ${\mathbb{F}}$.

 With this we can assure the fulfilment of the Lorentz condition in Equation
 \rf{Lorentzequation}.

The treatment goes on rewriting  explicitly all the equations \rf{N-1},\rf{N-2},\rf{N-3} :

\begin{eqnarray}
&&{a^{(-1)}}^{(\rho_N)}\xx\sum_{j=1}^N\Biggl[{\mathbb{G}}^{(\nu,\mu)}_{((\rho_1,\cdots,\rho_N),(\sigma_1,\cdots,\sigma_j))} \xx\Biggl(\cD_\nu\cD_\mu\Biggr) +{\mathbb{F}}^{(\nu,\mu)}_{((\rho_1,\cdots,\rho_N),
(\eta_1,\cdots,\eta_{k}))}\xx\nn\\&&
\cR^{(\eta_1,\cdots,\eta_n)}_{([\mu,\nu],\lambda)}\xx\delta^{(\lambda,\eta,_{n+1},\cdots,\eta_k)}_{(\sigma_1,\cdots,\sigma_j)}
+
{\mathbb{F}}^{(\nu,\mu)}
_{((\sigma_1,\cdots,\sigma_j),
(\eta_1,\cdots,\eta_{k}))}\xx\sum_{l=1}^k
\cR^{(\eta_1\cdots\eta_l)}_{([\mu,\nu],\lambda)}\xx\delta_{(\rho_1,\cdots,\rho_N)}^{(\lambda,\eta_{l+1},\cdots,\eta_k)}\Biggr]\cA^{(\sigma_1,\cdots,\sigma_j)}\xx\nn\\
&&=\sum_{j=1}^{N}\Biggl[{\mathbb{G}}^{(\nu,\mu)}_{((\rho_1,\cdots,\rho_{N-1}),(\sigma_1,\cdots,\sigma_j))} \xx\Biggl(\cD_\nu\cD_\mu\Biggr) +{\mathbb{F}}^{(\nu,\mu)}_{((\rho_1,\cdots,\rho_{N-1}),
(\eta_1,\cdots,\eta_{k}))}\xx\nn\\&&
\cR^{(\eta_1,\cdots,\eta_n)}_{([\mu,\nu],\lambda)}\xx\delta^{(\lambda,\eta,_{n+1},\cdots,\eta_k)}_{(\sigma_1,\cdots,\sigma_j)}
+
{\mathbb{F}}^{(\nu,\mu)}
_{((\sigma_1,\cdots,\sigma_j),
(\eta_1,\cdots,\eta_{k}))}\xx\sum_{l=1}^k
\cR^{(\eta_1\cdots\eta_l)}_{([\mu,\nu],\lambda)}\xx\delta_{(\rho_1,\cdots,\rho_{N-1})}^{(\lambda,\eta_{l+1},\cdots,\eta_k)}\Biggr]\cA^{(\sigma_1,\cdots,\sigma_j)}\xx\nn\\
\lbl{N-1A}
\end{eqnarray}

\begin{eqnarray}
&&{a^{(-1)}}^{(\rho_N,\rho_{N-1})}\xx\sum_{j=1}^N\Biggl[{\mathbb{G}}^{(\nu,\mu)}_{((\rho_1,\cdots,\rho_N),(\sigma_1,\cdots,\sigma_j))} \xx\Biggl(\cD_\nu\cD_\mu\Biggr)+\sum_{k=1}^j{\mathbb{F}}^{(\nu,\mu)}_{((\rho_1,\cdots,\rho_N),
(\eta_1,\cdots,\eta_{k}))}\xx\nn\\&&
\cR^{(\eta_1,\cdots,\eta_n)}_{([\mu,\nu],\lambda)}\xx\delta^{(\lambda,\eta,_{n+1},\cdots,\eta_k)}_{(\sigma_1,\cdots,\sigma_j)}
+
\sum_{k=1}^j{\mathbb{F}}^{(\nu,\mu)}
_{((\sigma_1,\cdots,\sigma_j),
(\eta_1,\cdots,\eta_{k}))}\xx\sum_{l=1}^k
\cR^{(\eta_1\cdots\eta_l)}_{([\mu,\nu],\lambda)}\xx\delta_{(\rho_1,\cdots,\rho_N)}^{(\lambda,\eta_{l+1},\cdots,\eta_k)}\Biggr]\cA^{(\sigma_1,\cdots,\sigma_j)}\xx\nn\\
&&=\sum_{j=1}^{N}\Biggl[{\mathbb{G}}^{(\nu,\mu)}_{((\rho_1,\cdots,\rho_{N-2}),(\sigma_1,\cdots,\sigma_j))} \xx\Biggl(\cD_\nu\cD_\mu\Biggr) +\sum_{k=1}^j{\mathbb{F}}^{(\nu,\mu)}_{((\rho_1,\cdots,\rho_{N-2}),
(\eta_1,\cdots,\eta_{k}))}\xx\nn\\&&
\cR^{(\eta_1,\cdots,\eta_n)}_{([\mu,\nu],\lambda)}\xx\delta^{(\lambda,\eta,_{n+1},\cdots,\eta_k)}_{(\sigma_1,\cdots,\sigma_j)}
+
\sum_{k=1}^j{\mathbb{F}}^{(\nu,\mu)}
_{((\sigma_1,\cdots,\sigma_j),
(\eta_1,\cdots,\eta_{k}))}\xx\sum_{l=1}^k
\cR^{(\eta_1\cdots\eta_l)}_{([\mu,\nu],\lambda)}\xx\delta_{(\rho_1,\cdots,\rho_{N-2})}^{(\lambda,\eta_{l+1},\cdots,\eta_k)}\Biggr]\cA^{(\sigma_1,\cdots,\sigma_j)}\xx\nn\\
\lbl{N-2A}
\end{eqnarray}

\begin{eqnarray}
&&{b^{(-1)}}^{(\rho_{N-2})}\xx\sum_{j=1}^{N-2}\Biggl[{\mathbb{G}}^{(\nu,\mu)}_{((\rho_1,\cdots,\rho_{N-2}),(\sigma_1,\cdots,\sigma_j))} \xx\Biggl(\cD_\nu\cD_\mu\Biggr) +\sum_{k=1}^j{\mathbb{F}}^{(\nu,\mu)}_{((\rho_1,\cdots,\rho_{N-2}),
(\eta_1,\cdots,\eta_{k}))}\xx\nn\\&&
\cR^{(\eta_1,\cdots,\eta_n)}_{([\mu,\nu],\lambda)}\xx\delta^{(\lambda,\eta,_{n+1},\cdots,\eta_k)}_{(\sigma_1,\cdots,\sigma_j)}
+
\sum_{k=1}^j{\mathbb{F}}^{(\nu,\mu)}
_{((\sigma_1,\cdots,\sigma_j),
(\eta_1,\cdots,\eta_{k}))}\xx\sum_{l=1}^k
\cR^{(\eta_1\cdots\eta_l)}_{([\mu,\nu],\lambda)}\xx\delta_{(\rho_1,\cdots,\rho_{N-2})}^{(\lambda,\eta_{l+1},\cdots,\eta_k)}\Biggr]\cA^{(\sigma_1,\cdots,\sigma_j)}\xx\nn\\
&&=\sum_{j=1}^{N}\Biggl[{\mathbb{G}}^{(\nu,\mu)}_{((\rho_1,\cdots,\rho_{N-3}),(\sigma_1,\cdots,\sigma_j))} \xx\Biggl(\cD_\nu\cD_\mu\Biggr) +\sum_{k=1}^j{\mathbb{F}}^{(\nu,\mu)}_{((\rho_1,\cdots,\rho_{N-3}),
(\eta_1,\cdots,\eta_{k}))}\xx\nn\\&&
\cR^{(\eta_1,\cdots,\eta_n)}_{([\mu,\nu],\lambda)}\xx\delta^{(\lambda,\eta,_{n+1},\cdots,\eta_k)}_{(\sigma_1,\cdots,\sigma_j)}
+
\sum_{k=1}^j{\mathbb{F}}^{(\nu,\mu)}
_{((\sigma_1,\cdots,\sigma_j),
(\eta_1,\cdots,\eta_{k}))}\xx\sum_{l=1}^k
\cR^{(\eta_1\cdots\eta_l)}_{([\mu,\nu],\lambda)}\xx\delta_{(\rho_1,\cdots,\rho_{N-3})}^{(\lambda,\eta_{l+1},\cdots,\eta_k)}\Biggr]\cA^{(\sigma_1,\cdots,\sigma_j)}\xx\nn\\
\lbl{N-3A}
\end{eqnarray}

  and selecting the various tensor fields content, from the double order covariant term monomials we obtain the conditions  for the functions  ${\mathbb{G}}$:
 \begin{eqnarray}
&&{\cV_{{10}}}^{(\nu,\mu)}_{((\rho_1,\cdots,\rho_{N-1}),(\sigma_1,\cdots,\sigma_j))} \xx 
\equiv {\mathbb{G}}^{(\nu,\mu)}_{((\rho_1,\cdots,\rho_{N-1}),(\sigma_1,\cdots,\sigma_j))} \xx-{{a^{}}^{-1}}^{(\rho_N)}\xx
{\mathbb{G}_i}^{(\nu,\mu)}_{((\rho_1,\cdots,\rho_N),(\sigma_1,\cdots,\sigma_j))} \xx=0
\nn\\
 &&{\cV_{11}}^{(\nu,\mu)}_{((\rho_1,\cdots,\rho_{N-2}),(\sigma_1,\cdots,\sigma_j))} \xx
 \equiv{\mathbb{G}}^{(\nu,\mu)}_{((\rho_1,\cdots,\rho_{N-2}),(\sigma_1,\cdots,\sigma_j))} \xx-{{a^{}}^{-1}}^{(\rho_N,\rho_{N-1})}\xx{\mathbb{G}}^{(\nu,\mu)}_{((\rho_1,\cdots,\rho_N),(\sigma_1,\cdots,\sigma_j))} \xx=0
\nn\\
 &&{\cV_{{12}}}^{(\nu,\mu)}_{((\rho_1,\cdots,\rho_{N-3}),(\sigma_1,\cdots,\sigma_j))} \xx
 \equiv{\mathbb{G}}^{(\nu,\mu)}_{((\rho_1,\cdots,\rho_{N-3}),(\sigma_1,\cdots,\sigma_j))} \xx-{{b^{}}^{-1}}^{(\rho_{N-2})}\xx{\mathbb{G}}^{(\nu,\mu)}_{((\rho_1,\cdots,\rho_{N-2}),(\sigma_1,\cdots,\sigma_j))} \xx=0
\nn\\
\nn\\
\lbl{symmetricconstraints}
\end{eqnarray}

{\it{Now a pleasant surprise is at our hands; since, if the previous conditions \rf{symmetricconstraints} are satisfied the links between the ${\mathbb{G}}$ and ${\mathbb{F}}$ terms, assured by Equation \rf{condtransv1},{\tt{ imply the complete fulfilment of the the ( remaining) ${\mathbb{F}}$ dependent part of Equations \rf{N-1A},\rf{N-2A},\rf{N-3A}  conditions}}.}} This welcome result indicates the quality of our treatment.

 At this stage we have to pick up from this Section, the $\cV$ conditions \rf{transversality}, \rf{ourF}, \rf{ourGn}, \rf{condtransv1}, \rf{symmetricconstraints}, and consider them as subsidiary conditions for a Lagrangian model.We remark that all the conditions can be settled in a way that their validity hold for any geometry.

\subsection{ A detailed example: the spin 2 model}
\lbl{spin2}
The most general Lagrangian, endowed with the auxiliary field tool kit, which, in our approach, would describe a massless spin -$2$ field, can be written:

\begin{eqnarray}
&&\cL= \int d^4 x \Biggl[ -\frac{1}{2} {\mathbb{G}}^{(\mu,\nu\}}_{((\rho_1,\rho_2),(\sigma_1,\sigma_2))}\xx\cD_{(\mu)}\cA^{(\rho_1,\rho_2)}\xx\cD_{(\nu)}\cA^{(\sigma_1,\sigma_2)}\xx+
-\frac{1}{2}{\mathbb{G}}^{(\mu,\nu)}_{((\rho_1),(\sigma_1,\sigma_2))}\xx
\cD_{(\mu)}\cA^{(\rho_1)}\xx\cD_{(\nu)}\cA^{(\sigma_1,\sigma_2)}\xx\nn\\
&&-\frac{1}{2}
{\mathbb{G}}^{(\mu,\nu)}_{((\rho_1,\rho_2),(\sigma_1))}\xx\cD_{(\mu)}\cA^{(\rho_1,\rho_2)}\xx\cD_{(\nu)}
\cA^{(\sigma_1,)}\xx-\frac{1}{2}{\mathbb{G}}^{(\mu,\nu)}_{((\rho_1),(\sigma_1))}\xx
\cD_{(\mu)}\cA^{(\rho_1)}\xx\cD_{(\nu)}\cA^{(\sigma_1)}\xx\nn\\
&&+{\mathbb{F}}^{(\mu,\nu)}_{((\rho_1,\rho_2),(\sigma_1,\sigma_2))}\xx
\cA^{(\rho_1,\rho_2)}\xx\cR^{(\sigma_1)}_{([\mu,\nu],\lambda)}\xx\cA^{(\lambda,\sigma_2)}\xx
+{\mathbb{F}}^{(\mu,\nu)}_{((\rho_1),(\sigma_1,\sigma_2))}\xx
\cA^{(\rho_1)}\xx\cR^{(\sigma_1)}_{([\mu,\nu],\lambda)}\xx\cA^{(\lambda,\sigma_2)}\xx\nn\\
&&
+{\mathbb{F}}^{(\mu,\nu)}_{((\rho_1,\rho_2),(\sigma_1,))}\xx
\cA^{(\rho_1,\rho_2)}\xx\cR^{(\sigma_1)}_{([\mu,\nu],\lambda)}\xx\cA^{(\lambda)}\xx
+{\mathbb{F}}^{(\mu,\nu)}_{((\rho_1),(\sigma_1))}\xx
\cA^{(\rho_1)}\xx\cR^{(\sigma_1)}_{([\mu,\nu],\lambda)}\xx\cA^{(\lambda)}\xx\Biggr]\nn\\
&&\equiv \int d^4 x \cL\xx\nn\\
\end{eqnarray}

This Lagrangian must be endowed, as discussed in the Section \rf{Introduction}, with the constraints which extract the highest spin $2$ content.
\begin{eqnarray}
&&\cD_{(\rho_1)}\cA^{(\rho_1,\rho_2)}\xx=0\nn\\
&&\cD_{(\rho)}\cA^{(\rho)}\xx=0\nn\\
&&a{(\rho_1)}\cA^{(\rho_1,\rho_2)}\xx=\cA^{(\rho_2)}\xx\nn\\
&&a{(\rho_1,\rho_2)}\cA^{(\rho_1,\rho_2)}\xx=0\nn\\
\end{eqnarray}

The fields would respect the symmetry transformations:
\begin{eqnarray}
\cS \cA^{(\rho_1,\rho_2)}\xx&=&\ccC^{\lambda}\cD_{\lambda}\cA^{(\rho_1,\rho_2)}\xx-
\cD_\lambda\ccC^{(\rho_1)}\xx \cA^{(\lambda,\rho_2)}\xx-\cD_\lambda \ccC^{(\rho_2)}\xx\cA^{(\rho_1,\lambda)}\xx\nn\\
\cS \cA^{(\rho)}\xx&=&\ccC^{\lambda}\cD_{\lambda}\cA^{(\rho)}\xx-\cD_\lambda\ccC^{(\rho)}\xx \cA^{(\lambda)}\xx\nn\\
\end{eqnarray}

The purpose of this appendix is to extract the (proper) highest spin equation of motion, according to the recipes described described in Section \rf{model}.In this case the chain of constraints is very short, so the analysis is straight and plausible.The treatment follows the lines of the previous section, finding the interesting result for the choice of the $a_{(\rho)}\xx$, $a_{(\rho_1,\rho_2)}\xx$ parameters that they must verify the consistency condition: ${a^{(-1)}}^{(\rho_1,\rho_2)}\xx{a}_{(\rho_2)}\xx=0$

This result put geometrical conditions on the process  to extract the highest field content.
We show now how this result comes out.
\vskip 1.cm

First of all the Equation of motion take the form:
\begin{eqnarray}
\Biggl[\frac{\delta \cS}{\delta \cA^{(\rho_1,\rho_2)}\xx}\Biggr]
&=&{\mathbb{G}}^{(\mu,\nu)}_{((\rho_1,\rho_2),(\sigma_1,\sigma_2))}\xx
\cD_{(\mu)}\cD_{(\mu)}\cA^{(\sigma_1,\sigma_2)}\xx+
{\mathbb{G}}^{(\mu,\nu)}_{((\rho_1,\rho_2),(\sigma_1))}\xx
\cD_{(\mu)}\cD_{(\mu)}\cA^{(\sigma_1)}\xx\nn\\
&+&{\mathbb{F}}^{(\mu,\nu)}_{((\rho_1,\rho_2),(\sigma_1,\sigma_2))}\xx
\cR^{(\sigma_1)}_{([\mu,\nu],\lambda)}\xx\cA^{(\lambda,\sigma_2)}\xx
+{\mathbb{F}}^{(\mu,\nu)}_{((\rho_1,\rho_2),(\sigma_1))}\xx
\cR^{(\sigma_1)}_{([\mu,\nu],\lambda)}\xx\cA^{(\lambda)}\xx\nn\\
&+&{\mathbb{F}}^{(\mu,\nu)}_{((\sigma_1,\sigma_2),(\eta_1,\eta_2))}\xx\cA^{(\sigma_1,\sigma_2)}\xx
\cR^{(\eta_1)}_{([\mu,\nu],\lambda)}\xx\delta^{(\lambda,\eta_2)}_{(\rho_1,\rho_2)}
+{\mathbb{F}}^{(\mu,\nu)}_{((\sigma_1),(\eta_1,\eta_2))}\xx\cA^{(\sigma_1)}\xx
\cR^{(\eta_1)}_{([\mu,\nu],\lambda)}\xx\delta^{(\lambda,\eta_2)}_{(\rho_1,\rho_2)}\nn\\
\lbl{motion2}
\end{eqnarray}

\begin{eqnarray}
\Biggl[\frac{\delta \cS}{\delta \cA^{(\rho_1)}\xx}\Biggr]&=&{\mathbb{G}}^{(\mu,\nu)}_{((\rho_1),(\sigma_1,\sigma_2))}\xx
\cD_{(\mu)}\cD_{(\nu)}\cA^{(\sigma_1,\sigma_2)}\xx+
{\mathbb{G}}^{(\mu,\nu)}_{((\rho_1),(\sigma_1))}\xx
\cD_{(\mu)}\cD_{(\nu)}\cA^{(\sigma_1)}\xx\nn\\
&+&{\mathbb{F}}^{(\mu,\nu)}_{((\rho_1),(\sigma_1,\sigma_2))}\xx
\cR^{(\sigma_1)}_{([\mu,\nu],\lambda)}\xx\cA^{(\lambda,\sigma_2)}\xx
+{\mathbb{F}}^{(\mu,\nu)}_{((\rho_1),(\sigma_1))}\xx
\cR^{(\sigma_1)}_{([\mu,\nu],\lambda)}\xx\cA^{(\lambda)}\xx\nn\\
&+&{\mathbb{F}}^{(\mu,\nu)}_{((\sigma_1,\sigma_2),(\eta_1))}\xx\cA^{(\sigma_1,\sigma_2)}\xx
\cR^{(\eta_1)}_{([\mu,\nu],\lambda)}\xx\delta^{(\lambda)}_{(\rho_1)}
+{\mathbb{F}}^{(\mu,\nu)}_{((\sigma_1),(\eta_1))}\xx\cA^{(\sigma_1)}\xx
\cR^{(\eta_1)}_{([\mu,\nu],\lambda)}\xx\delta^{(\lambda)}_{(\rho_1)}\nn\\
\lbl{motion1}
\end{eqnarray}

First of all we have to impose the (covariant) Lorentz conditions:

\begin{eqnarray}
\cD^{(\rho_1)}\Biggl[\frac{\delta \cS}{\delta \cA^{(\rho_1,\rho_2)}\xx}\Biggr]=0
\lbl{Lorentz2}
\end{eqnarray}
\begin{eqnarray}
\cD^{(\rho_1)}\Biggl[\frac{\delta \cS}{\delta \cA^{(\rho_1)}\xx}\Biggr]=0
\lbl{Lorentz1a}
\end{eqnarray}

Going straightforward to the results,
from Equation \rf{Lorentz2} we get:
\begin{eqnarray}
&&\cD^{(\rho_1)}{\mathbb{G}}^{(\mu,\nu)}_{((\rho_1,\rho_2),(\sigma_1,\sigma_2))}\xx=0\\
&&\cD^{(\rho_1)}{\mathbb{G}}^{(\mu,\nu)}_{((\rho_1,\rho_2),(\sigma_1))}\xx=0\nn\\
&&\cD^{(\rho_1)}{\mathbb{F}}^{(\mu,\nu)}_{((\rho_1,\rho_2),(\sigma_1,\sigma_2))}\xx=0\nn\\
&&\cD^{(\rho_1)}{\mathbb{F}}^{(\mu,\nu)}_{((\rho_1,\rho_2),(\sigma_1))}\xx=0\nn\\
&&\cD^{(\rho_1)}{\mathbb{F}}^{(\mu,\nu)}_{((\sigma_1,\sigma_2),(\eta_1,\eta_2))}\xx
\delta^{(\lambda,\eta_2)}_{(\rho_1,\rho_2)}=0\nn\\
&&\cD^{(\rho_1)}{\mathbb{F}}^{(\mu,\nu)}_{((\sigma_1),(\eta_1,\eta_2))}\xx
\delta^{(\lambda,\eta_2)}_{(\rho_1,\rho_2)}=0\nn\\
&&g^{(\rho_1,\lambda)}\xx
{\mathbb{G}}^{(\mu,\nu)}_{((\rho_1,\rho_2),(\sigma_1,\sigma_2))}\xx=\delta^{(\lambda)}_{(\sigma_2)}{\mathbb{H}}^{(\mu,\nu)}_{((\rho_1,\rho_2),(\sigma_1))}\xx\nn\\
\end{eqnarray}

and the conditions as the ones of the type of Equations \rf{condtransv1} are:
\begin{eqnarray}
&&g^{(\rho_1,\delta)}\xx\Biggl[\delta^{(\delta)}_{(\phi)}\biggl(
{\mathbb{G}}^{(\theta,\tau)}_{((\rho_1,\rho_2),(\omega,\sigma_2))}\xx\delta^{(\chi,\sigma_2)}_{(\xi_1,\xi_2)}
\biggr)\nn\\
&+&\delta^{(\tau)}_{(\phi)}\biggl({\mathbb{F}}^{([\delta,\theta])}_{((\rho_1,\rho_2),(\omega,\sigma_2))}\xx
\delta^{(\chi,\sigma_2)}_{(\xi_1,\xi_2)}+{\mathbb{F}}^{([\delta,\theta])}_{((\xi_1,\xi_2),(\omega,\eta_2))}\xx
\delta^{(\chi,\eta_2)}_{(\rho_1,\rho_2)}\biggr)\Biggr]\cR^{(\omega)}_{([\delta,\theta],\chi)}\xx
=0\nn\\
\lbl{spin2lorentz1}
\end{eqnarray}
\begin{eqnarray}
&&g^{(\rho_1,\phi)}\xx\Biggl[\delta^{(\delta)}_{(\phi)}\biggl(
{\mathbb{G}}^{(\theta,\tau)}_{((\rho_1,\rho_2),(\omega))}\xx\delta^{(\chi)}_{(\xi)}
\biggr)
\nn\\
&+&\delta^{(\tau)}_{(\phi)}\biggl(
{\mathbb{F}}^{(\delta,\theta)}_{((\rho_1,\rho_2),(\omega)))}\xx\delta^{(\chi)}_{(\xi)}+
{\mathbb{F}}^{(\delta,\theta)}_{((\xi),(\omega,\eta_1))}\xx\delta^{(\chi,\eta_1)}_{(\rho_1,\rho_2)}\biggr)\Biggr]
\cR^{(\omega)}_{([\delta,\theta],\chi)}\xx
=0\nn\\
\lbl{spin2lorentz2}
\end{eqnarray}
which link the role of the  ${\mathbb{G}}$ functions to the one of the ${\mathbb{F}}$'s.

While the Equations of the type of Equations \rf{A}
\begin{eqnarray}
&&g^{(\rho_1,\lambda)}\xx\Biggl[
{\mathbb{G}}^{(\mu,\nu)}_{((\rho_1,\rho_2),(\sigma_1,\sigma_2))}\xx
\Biggl(\biggl(\cD_{(\mu)}
\cR_{([\lambda,\nu]\eta)}^{\sigma_1}\xx\biggr)\delta^{(\eta,\sigma_2)}_{(\omega_1,\omega_2)}
+
\biggl(\cD_\mu\cR_{([\lambda,\nu]\eta)}^{\sigma_2}\xx\biggr)
\delta^{(\sigma_1,\eta)}_{(\omega_1,\omega_2)}\Biggr)\nn\\
&+&{\mathbb{F}}^{(\mu,\nu)}_{((\rho_1,\rho_2),(\sigma_1,\sigma_2))}\xx\biggl(
\cD_\lambda\cR^{(\sigma_1)}_{([\mu,\nu],\eta)}\xx\biggr)\delta^{(\eta,\sigma_2)}_{(\omega_1,\omega_2)}
+{\mathbb{F}}^{(\mu,\nu)}_{((\sigma_1,\sigma_2),(\eta_1,\eta_2))}\xx\delta^{(\sigma_1,\sigma_2)}_{(\omega_1,\omega_2)}\biggl(\cD_\lambda\cR^{(\eta_1)}_{([\mu,\nu],\tau)}\xx
\delta^{(\tau,\eta_2)}_{(\rho_1,\rho_2)}\biggr)\Biggr]=0\nn\\
\lbl{spin2A1}
\end{eqnarray}
\begin{eqnarray}
&&g^{(\rho_1,\lambda)}\xx\Biggl[
{\mathbb{G}}^{(\mu,\nu)}_{((\rho_1,\rho_2),(\sigma_1))}\xx
\biggl(\cD_\mu\cR_{([\lambda,\nu],\eta)}^{\sigma_1}\xx\biggr)\delta^{(\eta)}_{(\omega)}
+{\mathbb{F}}^{(\mu,\nu)}_{((\rho_1,\rho_2),(\sigma_1))}\xx
\biggl(\cD_\lambda\cR^{(\sigma_1)}_{([\mu,\nu],\eta)}\xx\biggr)\delta^{(\eta)}_{(\omega)}\nn\\
&+&{\mathbb{F}}^{(\mu,\nu)}_{((\sigma_1),(\eta_1,\eta_2))}\xx\Biggl(
\delta^{(\sigma_1)}_{(\omega)}
\biggl(\cD_\lambda\cR^{(\eta_1)}_{([\mu,\nu],\tau)}\xx\delta^{(\tau,\eta_2)}_{(\rho_1,\rho_2)}\biggr)\Biggr)\Biggr]=0
\nn\\
\lbl{spin2A2}
\end{eqnarray}
give cyclic conditions using the properties of the Bianchi equations.

Repeating the same procedure to Equation \rf{Lorentz1a} we derive:

\begin{eqnarray}
&&\cD^{(\rho_1)}{\mathbb{G}}^{(\mu,\nu)}_{((\rho_1),(\sigma_1,\sigma_2))}\xx=0\nn\\
&&\cD^{(\rho_1)}{\mathbb{G}}^{(\mu,\nu)}_{((\rho_1),(\sigma_1))}\xx=0\nn\\
&&\cD^{(\rho_1)}{\mathbb{F}}^{(\mu,\nu)}_{((\rho_1),(\sigma_1,\sigma_2))}\xx=0\nn\\
&&\cD^{(\rho_1)}{\mathbb{F}}^{(\mu,\nu)}_{((\rho_1),(\sigma_1))}\xx=0\nn\\
&&\cD^{(\rho_1)}{\mathbb{F}}^{(\mu,\nu)}_{((\sigma_1,\sigma_2),(\eta_1))}\xx
\delta^{(\lambda)}_{(\rho_1)}=0\nn\\
&&g^{(\rho_1,\lambda)}\xx
{\mathbb{G}}^{(\mu,\nu)}_{((\rho_1),(\sigma_1,\sigma_2))}\xx=\delta^{(\lambda)}_{(\sigma_2)}{\mathbb{H}}^{(\mu,\nu)}_{((\rho_1),(\sigma_1))}\xx\nn\\
\end{eqnarray}
and the constraints which have generalization in Equations \rf{condtransv1} are written:
\begin{eqnarray}
&&g^{(\rho_1,\phi)}\xx\Biggl[\delta^{(\delta)_{(\phi)}}\biggl({\mathbb{G}}^{(\theta,\tau)}_{((\rho_1),(\omega,\sigma_2))}\xx\delta^{(\chi,\sigma_2)}_{(\xi_1,\xi_2)}
\biggr)\nn\\
&+&\delta^{(\tau)}_{(\phi)}\biggl({\mathbb{F}}^{([\delta,\theta])}_{((\xi_1,\xi_2),(\omega))}\xx
\delta^{(\chi)}_{(\rho_1)}+{\mathbb{F}}^{([\delta,\theta])}_{((\rho_1),(\omega,\sigma_2))}\xx\delta^{(\chi,\sigma_2)}_{(\xi_1,\xi_2)}
\biggr)\Biggr]\cR^{(\omega)}_{([\delta,\theta],\chi)}\xx
=0\nn\\
\lbl{spin2lorentz3}
\end{eqnarray}
\begin{eqnarray}
&&g^{(\rho_1,\phi)}\xx\Biggl[\delta^{(\delta)}_{(\phi)}\biggl({\mathbb{G}}^{(\theta,\tau)}_{((\rho_1),(\omega))}\xx\delta^{(\chi)}_{(\xi)}
\biggr)\nn\\
&+&\delta^{(\tau)}_{(\phi)}\biggl({\mathbb{F}}^{([\delta,\theta])}_{((\rho_1),(\omega))}\xx\delta^{(\chi)}_{(\xi)}+
{\mathbb{F}}^{([\delta,\theta])}_{((\xi),(\omega))}\xx\delta^{(\chi)}_{(\rho_1)}\biggr)\Biggr]
\cR^{(\omega)}_{([\delta,\theta],\chi)}\xx
=0\nn\\
\lbl{spin2lorentz4}
\end{eqnarray}
while the ones which specialize for spin 2 the Equation \rf{A}:
\begin{eqnarray}
&&g^{(\rho_1,\lambda)}\xx\Biggl[
{\mathbb{G}}^{(\mu,\nu)}_{((\rho_1),(\sigma_1,\sigma_2))}\xx
\Biggl(\biggl(\cD_{(\mu)}
\cR_{([\lambda,\nu]\eta)}^{\sigma_1}\xx\biggr)\delta^{(\eta,\sigma_2)}_{(\omega_1,\omega_2)}
+\biggl(\cD_\mu\cR_{([\lambda,\nu]\eta)}^{\sigma_2}\xx\biggr)
\delta^{(\sigma_1,\eta)}_{(\omega_1,\omega_2)}\Biggr)
\nn\\
&+&
{\mathbb{F}}^{(\mu,\nu)}_{((\rho_1),(\sigma_1,\sigma_2))}\xx\biggl(
\cD_\lambda\cR^{(\sigma_1)}_{([\mu,\nu],\eta)}\xx\biggr)\delta^{(\eta,\sigma_2)}_{(\omega_1,\omega_2)}
+{\mathbb{F}}^{(\mu,\nu)}_{((\sigma_1,\sigma_2),(\eta_1))}\xx
\delta^{(\sigma_1,\sigma_2)}_{(\omega_1,\omega_2)}\biggl(\cD_\lambda\cR^{(\eta_1)}_{([\mu,\nu],\tau)}\xx
\delta^{(\tau)}_{(\rho_1)}\biggr)\Biggr]=0\nn\\
\lbl{spin2A3}
\end{eqnarray}

\begin{eqnarray}
&&g^{(\rho_1,\lambda)}\xx\Biggl[
{\mathbb{G}}^{(\mu,\nu)}_{((\rho_1),(\sigma_1))}\xx
\biggl(\cD_\mu\cR_{([\lambda,\nu],\eta)}^{\sigma_1}\xx\biggr)\delta^{(\eta)}_{(\omega)}
+{\mathbb{F}}^{(\mu,\nu)}_{((\rho_1),(\sigma_1))}\xx
\biggl(\cD_\lambda\cR^{(\sigma_1)}_{([\mu,\nu],\eta)}\xx\biggr)\delta^{(\eta)}_{(\omega)}\nn\\
&+&
{\mathbb{F}}^{(\mu,\nu)}_{((\sigma_1),(\eta_1))}\xx
\delta^{(\sigma_1)}_{(\omega)}
\biggl(\cD_\lambda\cR^{(\eta_1)}_{([\mu,\nu],\tau)}\xx\delta^{(\tau)}_{(\rho_1)}\biggr)\Biggr]=0\nn\\
\lbl{spin2A4}
\end{eqnarray}

Then if we put equal to zero the square bracket content of Equations  \rf{spin2lorentz1},\rf{spin2lorentz2}\rf{spin2lorentz3} \rf{spin2lorentz4}
 we derive conditions (which are valid for any geometry), which will find again in the following;
 and the Equations \rf{spin2A1}\rf{spin2A2}\rf{spin2A3}\rf{spin2A4})  give us  again the conditions in Equation \rf{cyclic} ( of Appendix
\rf{motionconstraints}), restricted to the cases $i,l=1,2$

 Anyhow  we want to show a peculiar aspect of the higher spin 2 selection process,
  due to the low spin content and, consequently, the shorter condition's chain to be prepared for its selection.

Introducing  the parameters ${a^{(-1)}}^{(\rho_2)}\xx$, ${a^{(-1)}}^{(\rho_1,\rho_2)}\xx$
we have to impose:
\begin{eqnarray}
{a^{(-1)}}^{(\rho_2)}\xx\Biggl[\frac{\delta \cS}{\delta \cA^{(\rho_1,\rho_2)}\xx}\Biggr]=\Biggl[\frac{\delta \cS}{\delta \cA^{(\rho_1)}\xx}\Biggr]\nn\\
\lbl{a1}
\end{eqnarray}
\begin{eqnarray}
{a^{(-1)}}^{(\rho_1,\rho_2)}\xx\Biggl[\frac{\delta \cS}{\delta \cA^{(\rho_1,\rho_2)}\xx}\Biggr]=0
\lbl{a2}
\end{eqnarray}

we define an inversion procedure such that:
\begin{eqnarray}
{a^{(-1)}}^{(\rho_2)}\xx {a}_{(\rho_1)}\xx=\delta^{(\rho_2)}_{(\rho_1)}
\end{eqnarray}

\begin{eqnarray}
{a^{(-1)}}^{(\rho_1,\rho_2)}\xx {a}_{(\sigma_1,\sigma_2)}\xx=\delta^{(\rho_,\rho_1)}_{(\sigma_1,\sigma_2)}
\end{eqnarray}

From the condition \rf{a1}we derive, isolating the independent terms from Equation \rf{motion2},\rf{motion1}:

\begin{eqnarray}
{a^{(-1)}}^{(\rho_2)}\xx{\mathbb{G}}^{(\mu,\nu)}_{((\rho_1,\rho_2),(\sigma_1,\sigma_2))}\xx&=&{\mathbb{G}}^{(\mu,\nu)}_{((\rho_1),(\sigma_1,\sigma_2))}\xx\nn\\
{a^{(-1)}}^{(\rho_2)}\xx{\mathbb{G}}^{(\mu,\nu)}_{((\rho_1,\rho_2),(\sigma_1))}\xx&=&{\mathbb{G}}^{(\mu,\nu)}_{((\rho_1),(\sigma_1))}\xx
\end{eqnarray}
so we desume:
\begin{eqnarray}
{\mathbb{G}}^{(\mu,\nu)}_{((\rho_1,\rho_2),(\sigma_1,\sigma_2))}\xx
&=&\frac{1}{2}\Biggr(a_{(\rho_2)}\xx{\mathbb{G}}^{(\mu,\nu)}_{((\rho_1),(\sigma_1,\sigma_2))}\xx+a_{(\rho_1)}\xx{\mathbb{G}}^{(\mu,\nu)}_{((\rho_2),(\sigma_1,\sigma_2))}\xx\Biggl)\nn\\
\lbl{G1}
\end{eqnarray}
\begin{eqnarray}
{\mathbb{G}}^{(\mu,\nu)}_{((\rho_1,\rho_2),(\sigma_1))}\xx&=&\frac{1}{2}\Biggr(a_{(\rho_2)}\xx{\mathbb{G}}^{(\mu,\nu)}_{((\rho_1),(\sigma_1))}\xx+a_{(\rho_1)}\xx{\mathbb{G}}^{(\mu,\nu)}_{((\rho_2),(\sigma_1))}\xx\Biggr)\nn\\
\lbl{G2}
\end{eqnarray}

Now, as shown in Section \rf{model}, we recall that the conditions \rf{Lorentz2}, \rf{Lorentz1a}, together with Equations \rf{G1},\rf{G2}, completely solve the problems for the ${\mathbb{F}}$ too.
Anyhow, to reinforce the result we obtain in the following, we proceed in a naive way continuing the analisys of condition \rf{a1}
obtaining:

\begin{eqnarray}
&&{a^{(-1)}}^{(\rho_2)}\xx\Biggl[{\mathbb{F}}^{(\delta,\theta)}_{((\rho_1,\rho_2),(\omega,\sigma_2))}\xx\delta^{(\chi,\sigma_2)}_{(\kappa_1,\kappa_2)}+ {\mathbb{F}}^{([\delta,\theta])}_{((\rho_1,\rho_2),(\sigma_1,\omega))}\xx\delta^{(\sigma_1,\chi)}_{(\kappa_1,\kappa_2)}
\Biggr]
\nn\\
&=&\Biggl[{\mathbb{F}}^{([\delta,\theta])}_{((\rho_1),(\omega,\sigma_2))}\xx
\delta^{(\chi,\sigma_2)}_{(\kappa_1,\kappa_2)}+
{\mathbb{F}}^{([\delta,\theta])}_{((\kappa_1,\kappa_2),(\omega))}\xx
\delta^{(\chi)}_{(\rho_1)}\Biggr]
\nn\\
\end{eqnarray}

\begin{eqnarray}
&&{a^{(-1)}}^{(\rho_2)}\xx\Biggl[{\mathbb{F}}^{([\delta,\theta])}_{((\rho_1,\rho_2),(\omega))}\xx\delta^{(\chi)}_{(\kappa)}
+{\mathbb{F}}^{([\delta,\theta])}_{((\kappa),(\omega,\eta_2))}\xx\delta^{(\chi,\eta_2)}_{(\rho_1,\rho_2)}\Biggr]
\nn\\
&=&\Biggl[{\mathbb{F}}^{([\delta,\theta])}_{((\rho_1),(\omega))}\xx\delta^{(\chi)}_{(\kappa)}+
{\mathbb{F}}^{([\delta,\theta])}_{((\kappa),(\omega))}\xx\delta^{(\chi)}_{(\rho_1)}
\Biggr]
\nn\\
\end{eqnarray}

So we derive::

\begin{eqnarray}
&&\Biggl[{\mathbb{F}}^{([\delta,\theta])}_{((\rho_1,\rho_2),(\omega))}\xx\delta^{(\chi)}_{(\kappa)}
+{\mathbb{F}}^{([\delta,\theta])}_{((\kappa),(\omega,\eta_2))}\xx\delta^{(\chi,\eta_2)}_{(\rho_1,\rho_2)}\Biggr]
\nn\\
&=&\frac{1}{2}\Biggl({a}_{(\rho_2)}\xx\Biggl[{\mathbb{F}}^{([\delta,\theta])}_{((\rho_1),(\omega))}\xx\delta^{(\chi)}_{(\kappa)}
+{\mathbb{F}}^{([\delta,\theta])}_{((\kappa),(\omega))}\xx\delta^{(\chi)}_{(\rho_1)}
\Biggr]+{a}_{(\rho_1)}\xx\Biggl[{\mathbb{F}}^{([\delta,\theta])}_{((\rho_2),(\omega))}\xx\delta^{(\chi)}_{(\kappa)}
+{\mathbb{F}}^{([\delta,\theta])}_{((\kappa),(\omega))}\xx\delta^{(\chi)}_{(\rho_2)}
\Biggr]\Biggr)
\nn\\
\lbl{F1}
\end{eqnarray}

\begin{eqnarray}
&&\Biggl[{\mathbb{F}}^{([\delta,\theta])}_{((\rho_1,\rho_2),(\omega,\sigma_2))}\xx\delta^{(\chi,\sigma_2)}_{(\kappa_1,\kappa_2)}+ {\mathbb{F}}^{([\delta,\theta])}_{((\kappa_1,\kappa_2),(\sigma_1,\omega))}\xx
\delta^{(\sigma_1,\chi)}_{(\rho_1,\rho_2)}
\Biggr]
\nn\\
&=&\frac{1}{2}\Biggl({a}_{(\rho_2)}\xx\Biggl[{\mathbb{F}}^{([\delta,\theta])}_{((\rho_1),(\omega,\sigma_2))}\xx
\delta^{(\chi,\sigma_2)}_{(\kappa_1,\kappa_2)}\nn\\
&+&
{\mathbb{F}}^{([\delta,\theta])}_{((\kappa_1,\kappa_2),(\omega))}\xx
\delta^{(\chi)}_{(\rho_1)}\Biggr]+{a}_{(\rho_1)}\xx\Biggl[{\mathbb{F}}^{([\delta,\theta])}_{((\rho_2),(\omega,\sigma_2))}\xx
\delta^{(\chi,\sigma_2)}_{(\kappa_1,\kappa_2)}+
{\mathbb{F}}^{([\delta,\theta])}_{((\kappa_1,\kappa_2),(\omega))}\xx
\delta^{(\chi)}_{(\rho_2)}\Biggr]\Biggr)
\nn\\
\lbl{F2}
\end{eqnarray}

Going now to condition  \rf{a2} applied to the equation of motion \rf{motion2},\rf{motion1} isolating the second derivative term, we derive

\begin{eqnarray}
&&{a^{(-1)}}^{(\rho_1,\rho_2)}\xx
{\mathbb{G}}^{(\mu,\nu)}_{((\rho_1,\rho_2),(\sigma_1,\sigma_2))}\xx=0\nn\\
&&{a^{(-1)}}^{(\rho_1,\rho_2)}\xx{\mathbb{G}}^{(\mu,\nu)}_{((\rho_1,\rho_2),(\sigma_1))}\xx=0\nn\\
\lbl{ker1}
\end{eqnarray}

\begin{eqnarray}
&&{a^{(-1)}}^{(\rho_1,\rho_2)}\xx\Biggl[
{\mathbb{F}}^{(\delta,\theta)}_{((\rho_1,\rho_2),(\omega,\sigma_2))}\xx\delta^{(\chi,\sigma_2)}_{(\kappa_1,\kappa_2)}+
{\mathbb{F}}^{(\delta,\theta)}_{((\kappa_1,\kappa_2),(\omega,\eta_2))}
\delta^{(\chi,\eta_2)}_{(\rho_1,\rho_2)}\Biggr]
=0\nn\\
\end{eqnarray}

anf from the non-derivated terms we get:

\begin{eqnarray}
&&{a^{(-1)}}^{(\rho_1,\rho_2)}\xx\Biggl[{\mathbb{F}}^{([\delta,\theta])}_{((\rho_1,\rho_2),(\omega))}\xx\delta^{(\chi)}_{(\kappa)}+
{\mathbb{F}}^{([\delta,\theta])}_{((\kappa),(\omega,\eta_2))}\xx\delta^{(\chi,\eta_2)}_{(\rho_1,\rho_2)}\Biggr]
=0\nn\\
\end{eqnarray}

Collecting together all these results, we must check the consistency
 with Equations \rf{G1},\rf{G2},\rf{F1},\rf{F2}.
The full validity is verified if:

\begin{eqnarray}
{a^{(-1)}}^{(\rho_1,\rho_2)}\xx{\mathbb{G}}^{(\mu,\nu)}_{((\rho_1,\rho_2),(\sigma_1,\sigma_2))}\xx&=&\frac{1}{2}{a^{(-1)}}^{(\rho_1,\rho_2)}\xx \Biggl( a_{(\rho_2)}\xx{\mathbb{G}}^{(\mu,\nu)}_{((\rho_1),(\sigma_1,\sigma_2))}\xx
+a_{(\rho_1)}\xx{\mathbb{G}}^{(\mu,\nu)}_{((\rho_2),(\sigma_1,\sigma_2))}\xx\Biggr)
\nn\\&=&{a^{(-1)}}^{(\rho_1,\rho_2)}\xx a_{(\rho_1)}\xx{\mathbb{G}}^{(\mu,\nu)}_{((\rho_2),(\sigma_1,\sigma_2))}\xx=0\nn\\
{a^{(-1)}}^{(\rho_1,\rho_2)}\xx{\mathbb{G}}^{(\mu,\nu)}_{((\rho_1,\rho_2),(\sigma_1))}\xx&=&\frac{1}{2}
{a^{(-1)}}^{(\rho_1,\rho_2)}\xx\Biggl( a_{(\rho_2)}\xx{\mathbb{G}}^{(\mu,\nu)}_{((\rho_1),(\sigma_1))}\xx+ a_{(\rho_1)}\xx{\mathbb{G}}^{(\mu,\nu)}_{((\rho_2),(\sigma_1))}\xx\Biggr)\nn\\
&=&{a^{(-1)}}^{(\rho_1,\rho_2)}\xx a_{(\rho_1)}\xx{\mathbb{G}}^{(\mu,\nu)}_{((\rho_2),(\sigma_1))}\xx=0
\lbl{ker2}
\end{eqnarray}
(for the ${\mathbb{G}}$ functions)

while  for the ${\mathbb{F}}$'s we get:

\begin{eqnarray}
&&{a^{(-1)}}^{(\rho_1,\rho_2)}\xx\Biggl[
\frac{1}{2}\Biggl({a}_{(\rho_2)}\xx\Biggl[{\mathbb{F}}^{([\delta,\theta])}_{((\rho_1),(\omega,\sigma_2))}\xx
\delta^{(\chi,\sigma_2)}_{(\kappa_1,\kappa_2)}+
{\mathbb{F}}^{([\delta,\theta])}_{((\kappa_1,\kappa_2),(\omega))}\xx
\delta^{(\chi)}_{(\rho_1)}\Biggr]\nn\\
&&+{a}_{(\rho_1)}\xx\Biggl[{\mathbb{F}}^{([\delta,\theta])}_{((\rho_2),(\omega,\sigma_2))}\xx
\delta^{(\chi,\sigma_2)}_{(\kappa_1,\kappa_2)}+
{\mathbb{F}}^{([\delta,\theta])}_{((\kappa_1,\kappa_2),(\omega))}\xx
\delta^{(\chi)}_{(\rho_2)}\Biggr]\Biggr)
\Biggr]\nn\\
&&={a^{(-1)}}^{(\rho_1,\rho_2)}\xx{a}_{(\rho_1)}\xx\Biggl[{\mathbb{F}}^{([\delta,\theta])}_{((\rho_2),(\omega,\sigma_2))}\xx
\delta^{(\chi,\sigma_2)}_{(\kappa_1,\kappa_2)}+
{\mathbb{F}}^{([\delta,\theta])}_{((\kappa_1,\kappa_2),(\omega))}\xx
\delta^{(\chi)}_{(\rho_2)}\Biggr]
=0\nn\\
\lbl{ker3}
\end{eqnarray}

\begin{eqnarray}
&&{a^{(-1)}}^{(\rho_1,\rho_2)}\xx\Biggl[
\frac{1}{2}\Biggl({a}_{(\rho_2)}\xx\Biggl[{\mathbb{F}}^{([\delta,\theta])}_{((\rho_1),(\omega))}\xx\delta^{(\chi)}_{(\kappa)}
+{\mathbb{F}}^{([\delta,\theta])}_{((\kappa),(\omega))}\xx\delta^{(\chi)}_{(\rho_1)}
\Biggr]\nn\\
&&+{a}_{(\rho_1)}\xx\Biggl[{\mathbb{F}}^{([\delta,\theta])}_{((\rho_2),(\omega))}\xx\delta^{(\chi)}_{(\kappa)}
+{\mathbb{F}}^{([\delta,\theta])}_{((\kappa),(\omega))}\xx\delta^{(\chi)}_{(\rho_2)}
\Biggr]\Biggr)
\Biggr]\nn\\
&&={a^{(-1)}}^{(\rho_1,\rho_2)}\xx{a}_{(\rho_1)}\xx\Biggl[{\mathbb{F}}^{([\delta,\theta])}_{((\rho_2),(\omega))}\xx\delta^{(\chi)}_{(\kappa)}
+{\mathbb{F}}^{([\delta,\theta])}_{((\kappa),(\omega))}\xx\delta^{(\chi)}_{(\rho_2)}
\Biggr]
=0\nn\\
\lbl{ker4}
\end{eqnarray}
 So from \rf{ker1},\rf{ker2},\rf{ker3},\rf{ker4}   we recover the interesting (and reinforced) result:

\begin{eqnarray}
{a^{(-1)}}^{(\rho_1,\rho_2)}\xx{a}_{(\rho_2)}\xx=0
\end{eqnarray}
which put a geometrical constraint for the fields sectioning which extract the highest spin $2$ content.
It is obvious that this condition is valid only for this case.
\skip 1.0 cm.

The finishing touch of this part requires the inclusion of all the previous constraints in the (gauge model) machinery previously seen, with the conclusions  contained (as particular cases) in Section \rf{model}.

\end{document}